\documentclass[10pt,conference]{IEEEtran}
\IEEEoverridecommandlockouts
\usepackage{cite}
\usepackage{amsmath,amssymb,amsfonts}
\usepackage{algorithmic}
\usepackage{graphicx}
\usepackage{textcomp}
\usepackage{xcolor}
\usepackage[hyphens]{url}
\usepackage{fancyhdr}
\usepackage{hyperref}
\usepackage{xspace}
\usepackage{enumitem}
\usepackage{booktabs}
\usepackage{makecell}
\usepackage{eso-pic}

\title{\fontsize{19}{20}\selectfont\papername: Efficiently Using Large-Granularity Memory Systems \\ with Transparent Two-Dimensional Memory Layout\vspace{-0.2in}}

\author{
\IEEEauthorblockN{
Jae Hyung Ju\textsuperscript{*},
Euijun Chung\textsuperscript{*},
Hritvik Taneja,
Anish Saxena,
Shinnung Jeong,
Hyesoon Kim,
Moinuddin Qureshi
}

\IEEEauthorblockA{
\emph{
\{jhju, euijun, htaneja3, asaxena317, sjeong306, moin\}@gatech.edu, hyesoon@cc.gatech.edu
}
}
\IEEEauthorblockA{
Georgia Institute of Technology\vspace{-0.2in}
}
\thanks{\textsuperscript{*}Equal contribution.}
}

\begin{document}

\newcommand{\papername}{TileLens\xspace}

\newif\ifshowcomments
\showcommentsfalse

\ifshowcomments
  \newcommand\jcomment[1]{\noindent{\color{magenta}{\bf \fbox{J}}{\bf \it#1}}}
  \newcommand\ecomment[1]{\noindent{\color{blue}{\bf \fbox{E}}{\bf \it#1}}}
  \newcommand\fixme[1]{\noindent{\color{purple}{\bf \fbox{FIXME}}{\bf \it#1}}}
  \newcommand\moin[1]{\noindent{\color{red} {\bf \fbox{Moin}}{\bf \it#1}}}
\else
  \newcommand\jcomment[1]{}
  \newcommand\ecomment[1]{}
  \newcommand\fixme[1]{}
  \newcommand\moin[1]{}
\fi

\newif\ifshowrev
\showrevfalse 
\newcommand\rev[1]{%
  \ifshowrev
    {\color{blue}#1}%
  \else
    #1%
  \fi
}

\newcommand\tbd[1]{{\color{red}[#1]}}

\newcommand{\parhead}[1]{\vspace{0.1em}\noindent\textbf{#1}}

\newcommand{\parheadd}[1]{\vspace{0.55em}\noindent\textbf{#1}}
\pagenumbering{arabic}

\maketitle

\ifdefined\hpcacameraready 
  \thispagestyle{camerareadyfirstpage}
  \pagestyle{empty}
\else
  \thispagestyle{plain}
  \pagestyle{plain}
\fi

\AddToShipoutPictureBG*{%
  \AtPageUpperLeft{\raisebox{-0.5in}{\makebox[\paperwidth]{\small
    To appear in 2027 IEEE International Symposium on High-Performance Computer Architecture (HPCA)}}}}

\newcommand{\hpcaheight}{0mm}
\ifdefined\eaopen
\renewcommand{\hpcaheight}{12mm}
\fi
  

\begin{abstract}

Large Language Model (LLM) inference is bottlenecked by the capacity and bandwidth of GPU \emph{High-Bandwidth Memory (HBM)}. Recent proposals, such as \emph{High-Bandwidth Flash (HBF)} and \emph{RoMe}, offer higher capacity or bandwidth than HBM, but require a minimum access granularity of kilobytes.
We show that these \emph{Large-Granularity Memory Systems (LGMS)} can degrade LLM inference performance by up to an order of magnitude. The root cause is \emph{read amplification} during tiled matrix multiplication, in which each memory request fetches far more data than the compute tile needs. Such waste stems from a fundamental mismatch between the two-dimensional nature of compute tiles and the one-dimensional memory layout, causing each request to spill well beyond tile boundaries.

To eliminate read amplification in LGMS, we propose to use \emph{tile-major layout}. Rather than storing data as a one-dimensional strip, tile-major layout reshapes each contiguous memory block into a two-dimensional rectangle, aligning memory granularity with tile boundaries.
To ease the adoption of tile-major layout on GPUs, we propose \emph{TileLens}, lightweight software and hardware extensions that cover major classes of GPU kernels.
\emph{TileLens-SW} extends GPU DSLs so that DSL-based kernels can adopt tile-major by only changing the layout descriptor.
\emph{TileLens-HW} extends the Tensor Memory Accelerator (TMA) for tile-major support in TMA-based kernels without code changes. 
We evaluate TileLens on a cycle-level GPU simulator using Qwen-3 30B and Llama-3.1 70B.
Combining a tile-major layout with our adaptive degree prefetcher, \emph{TileLens} achieves near-HBM performance on HBF-augmented GPUs with a 5\,$\mu$s NAND read latency, reducing the geomean slowdown from 1.62--6.69$\times$ with conventional layouts to within 10\% of an HBM-only baseline.

\end{abstract}

\section{Introduction}
\label{sec:intro}

The rapid adoption of Large Language Models (LLMs) is pushing the limits of current GPU memory systems~\cite{ma2026challenges}.
Model sizes have grown to terabytes of memory~\cite{shoeybi2019megatron, achiam2023gpt, grattafiori2024llama, liu2024deepseek}, and emerging capabilities such as reasoning, multimodality, and agentic behavior require a larger per-user key-value (KV) cache.
As a result, GPU memory capacity and bandwidth have become the main bottlenecks of LLM inference performance~\cite{ma2026challenges}.
New GPU memory technologies have been proposed to address these bottlenecks by providing higher capacity or bandwidth than the current High-Bandwidth Memory (HBM).
\emph{High-Bandwidth Flash} (HBF)~\cite{ma2026challenges, sandisk2025hbf, ha2026h, hsu2026haven} is an emerging memory technology that stacks NAND flash dies in a way similar to HBM, offering 8--16$\times$ the capacity with comparable bandwidth.
\emph{RoMe}~\cite{nam2026rome} enlarges the access granularity of HBM so that the address and command buses can be repurposed for data transfer, increasing bandwidth.
Both proposals share a common architectural consequence: a minimum access granularity of kilobytes, far larger than HBM's 32\,B.
However, the impact of employing such \emph{Large-Granularity Memory Systems} (LGMS) on GPUs for LLM inference has not been well studied.

\begin{figure*}[t!]
    \centering
\vspace{-0.1in}
    \includegraphics[width=0.95\linewidth]{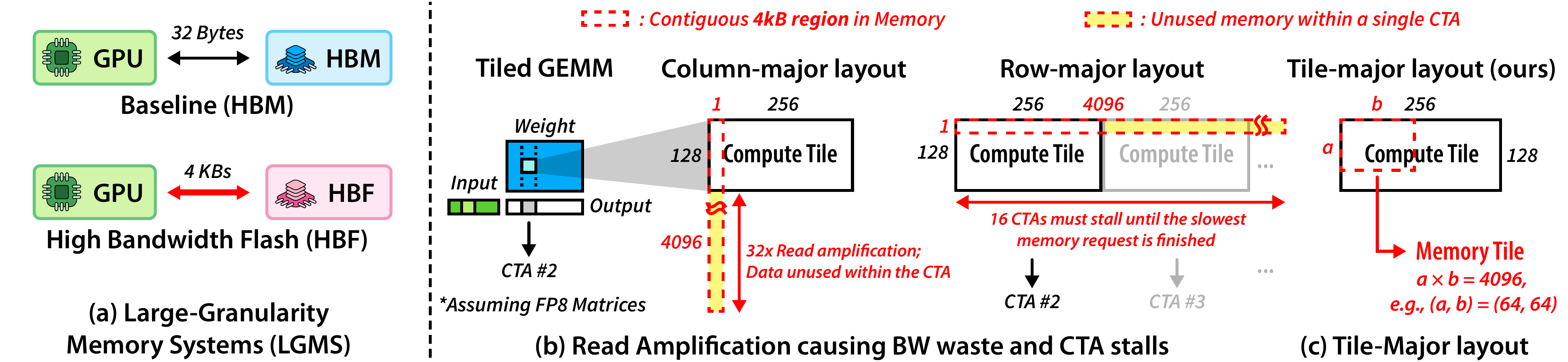}
   \vspace{-0.08in}
    \caption{(a) HBF and RoMe both require a minimum access granularity of 4\,KB, 128$\times$ coarser than HBM's 32\,B sectors. (b) Under 1-D (row- or column-major) memory layout, each 4\,KB access fetches data mostly outside the compute tile, wasting bandwidth and stalling CTAs that await the same memory request. (c) Tile-major layout shapes the 4\,KB region as a 2-D rectangle (e.g., $64 \times 64$ for FP8) that fits within the compute tile, such that every fetched byte is consumed.}
    \label{fig:intro_main}
\end{figure*}

We show that using LGMS on GPUs results in \emph{read amplification}, where each memory access fetches more data than the computation needs.
The dominant kernels in LLM inference are matrix multiplications (matmuls), which split matrices into two-dimensional (2-D) \emph{compute tiles} and schedule them across CTAs (cooperative thread arrays, or thread blocks)~\cite{nvidia2026cublas, nvidia2026cutlass, tillet2019triton}.
The width of a compute tile is typically 64--512 bytes, much smaller than the kilobyte-scale granularity of LGMS.
Thus, each LGMS access fetches a one-dimensional strip that extends far beyond the two-dimensional compute tile, as shown in Figure~\ref{fig:intro_main}-(b).

Read amplification degrades GPU performance by up to an order of magnitude with conventional layouts.
When weight matrices are stored in column- or row-major layout, an entire column or row is contiguous in memory.
In column-major layout, the excess data fetched by each CTA is not used by any other CTA, so read amplification directly wastes bandwidth.
Caching the excess data is infeasible: on an H200 GPU with $32\times$ amplification, hundreds of megabytes must be buffered to avoid the waste.
On an HBF-augmented GPU, column-major layout that suffers $16\times$ read amplification can degrade performance by $11.2\times$ (Section~\ref{sec:motivation}).

In row-major layout, overfetched data may be reused by other CTAs, but this sharing introduces a \emph{straggler} problem.
Because CTAs now share memory requests, no CTA can proceed until all shared requests arrive. 
This causes the slowest request to block every dependent CTA, stalling the execution of multiple CTAs.
We show that the straggler problem from read amplification also degrades execution time by $3.3\times$ on HBF-augmented GPUs.

Our insight is that if the contiguous memory block is two-dimensional and fits within a compute tile, read amplification will be eliminated.
Therefore, the \emph{goal} of this paper is to design low-cost architectural support for GPUs that maintains the benefits of LGMS without read amplification.
To this end, we propose to use \emph{tile-major layout} for LGMS.
As shown in Figure~\ref{fig:intro_main}-(c), tile-major layout reshapes the memory contiguity from a 1-D strip to a 2-D rectangle.
Since the rectangular tiles consumed by CTAs are now contiguous in global memory, tile-major layout greatly reduces the read amplification in tiled matmul kernels.

However, adopting tile-major layout is not straightforward, because tile-major is not a single layout but a family of layouts parameterized by the memory tile dimensions.
Since the correctness of a GPU kernel depends on the global memory layout it assumes, each memory tile shape requires its own kernel code.
To mitigate the effort to rewrite kernels for each shape, we propose \emph{\papername}, lightweight software and hardware extensions that let a single kernel transparently support any memory tile configuration.

\emph{\papername-SW} extends GPU DSLs (Domain-Specific Languages) to support tile-major in global memory.
While the layout abstraction of existing GPU DSLs can represent arbitrary layouts, kernels cannot directly adopt tile-major if their code assumes 1-D global memory layouts.
This assumption was never a problem: HBM's fine access granularity lets simple 1-D layouts already achieve high bandwidth utilization.
\papername-SW enables tile-major by viewing a tiled \mbox{2-D} matrix as a 4-D tensor while remaining compatible with existing kernel code.
Through this change, kernel libraries (e.g., CUTLASS~\cite{nvidia2026cutlass}) and high-performance kernels (e.g., FlashAttention~\cite{dao2023flashattention}) built on DSL frameworks can adopt tile-major layout by changing only the DSL layout descriptor.

\emph{\papername-HW} provides transparent tile-major layout support for kernels that do not use DSLs.
Our key insight is that modern GPUs already ship dedicated hardware that converts logical tensor indices into memory addresses, such as the NVIDIA Tensor Memory Accelerator (TMA)~\cite{nvidia2024h100}, AMD shim DMA~\cite{amd2026am020}, and Intel 2-D block loads~\cite{intel2025simd}.
Since the index-to-address mapping is fully contained in this hardware, it is a natural place to apply tile-major remapping.
\papername-HW minimally extends the TMA, enabling existing kernels and binaries that assume \mbox{1-D} layouts to support multiple memory tile configurations without rewriting code or recompiling.

While tile-major layout addresses read amplification on LGMS, integrating HBF into a GPU system requires additional microarchitectural changes.
To address HBF's limited write endurance, we allocate only weight matrices to HBF while keeping the KV cache and activations on HBM.
For HBF to coexist with HBM, we propose a mixed-granularity L2 cache and 4\,KB-granularity miss status holding registers (MSHRs) that track more concurrent HBF requests.
Finally, an adaptive degree prefetcher keeps enough requests in flight to hide the microsecond-scale HBF read latency, which is orders of magnitude longer than HBM's.
We evaluate TileLens on a cycle-level simulator and show that tile-major layout brings HBF-augmented GPU performance from a 1.62--6.69$\times$ geomean slowdown to within 10\% of an HBM-only system.

\parhead{Contribution.} This paper makes the following contributions.
\begin{enumerate}
    \item To the best of our knowledge, this is the first paper to analyze the read amplification problem on GPUs with LGMS. We show that read amplification degrades GPU performance by an order of magnitude, through both bandwidth waste and straggler-induced stalls.
    \item We propose using \emph{tile-major layout} for LGMS, where a contiguous memory region is mapped to 2-D \emph{memory tiles}, eliminating read amplification.
    \item We propose \emph{\papername}, which extends either GPU DSLs \emph{(\papername-SW)} or TMA \emph{(\papername-HW)} to transparently support tile-major layouts with various shapes and sizes.
    \item We propose an adaptive degree prefetcher that hides $\mu$s-scale HBF latency, and show on Qwen-3 30B~\cite{yang2025qwen3} and Llama-3.1 70B~\cite{grattafiori2024llama} that \papername reduces a 1.62--6.69$\times$ slowdown to within 10\% of an HBM baseline.
\end{enumerate}

\section{Background and Motivation}
\label{sec:background}

\subsection{Memory Capacity and Bandwidth Wall in LLM Inference}
\label{sec:bg_llm}

LLM inference consists of a prefill and a decode phase, and the decode phase dominates end-to-end latency as reasoning workloads produce increasingly long outputs~\cite{guo2025deepseek}.
Each decode step must read the entire model weights and the per-user KV cache to generate a single token, so each weight is used only once per token; this low arithmetic intensity makes decode latency bound by memory bandwidth.
Batching amortizes the weight reads across tokens to improve throughput, but each added user also grows the KV cache, so memory capacity caps further throughput gains.

\subsection{Emerging Memory Systems for GPUs}
\label{sec:bg_memory}

Several memory systems have been proposed to overcome the capacity and bandwidth limitations of HBM for LLM inference.
These systems share a common architectural consequence: a minimum read granularity of kilobytes (e.g., 4\,KB), far larger than HBM's 32\,B.
We refer to them collectively as \emph{Large-Granularity Memory Systems} (LGMS).

\parhead{HBF.}
\label{sec:bg_hbf} High-Bandwidth Flash (HBF)~\cite{sandisk2025hbf} stacks NAND flash dies, co-packaged with GPUs via a silicon interposer.
HBF provides bandwidth comparable to HBM (${\sim}$1.6\,TB/s per stack for Gen1) while offering 8--16$\times$ the capacity per stack due to the higher density of NAND flash cells.
Its access granularity is on the order of kilobytes, constrained by NAND page size and ECC, and its read latency is on the order of microseconds (1--10\,$\mu$s), against HBM's ${\sim}$100\,ns.

\parhead{RoMe.}
\label{sec:bg_rome} RoMe~\cite{nam2026rome} increases the access granularity of HBM to 4\,KB.
Fewer commands per unit of data free up address pins to carry data, increasing data bandwidth by 12.5\%.
By amortizing each command over a 4\,KB transfer, RoMe reduces command overhead by $128\times$ and reclaims that bandwidth for data.
While RoMe targets bandwidth rather than capacity, it arrives at the same 4\,KB access granularity as HBF, so the read amplification problem we identify applies equally to both.

\subsection{Matrix Multiplication on GPUs}
\begin{figure}[h]
    \centering
    \includegraphics[width=0.9\linewidth]{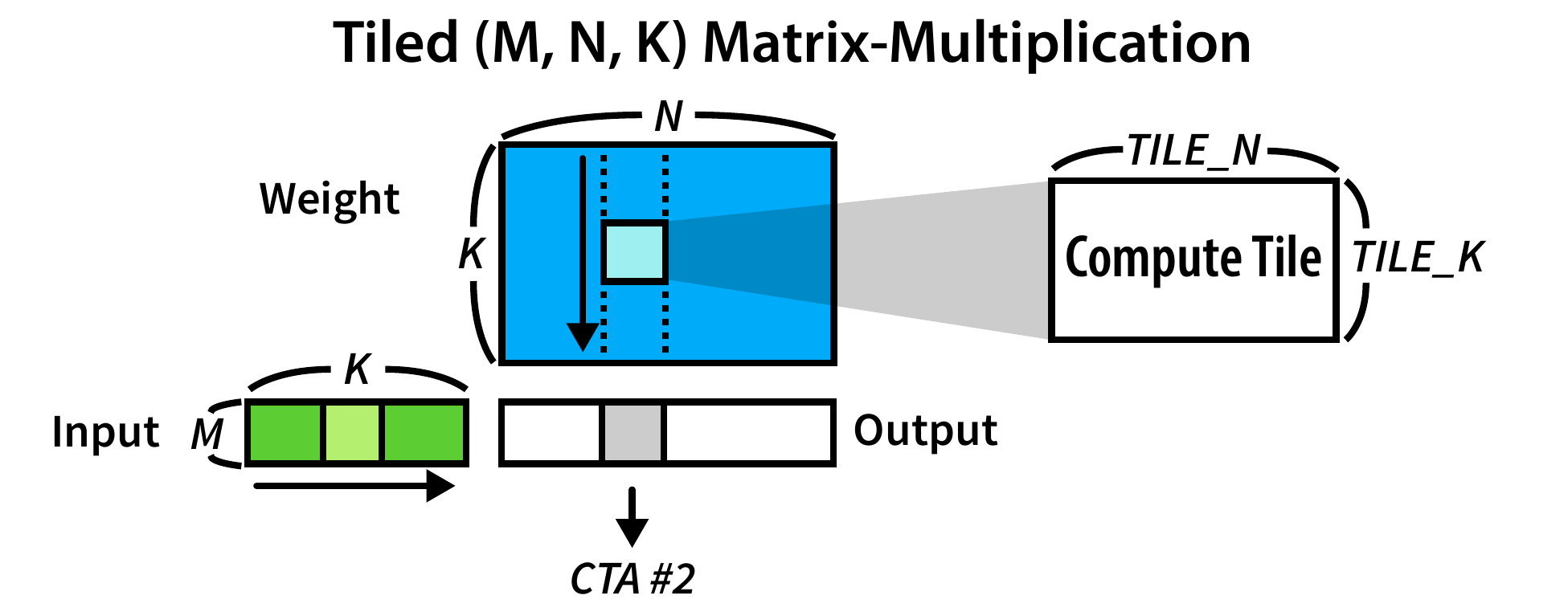}
    \vspace{-0.15in}
    \caption{Tiled matrix multiplication. Each CTA loads a compute tile from global memory and reduces along the K dimension to calculate the output.}
    \label{fig:gemm}
\end{figure}

Most of the time in LLM decode is spent on matrix multiplications, where for an input matrix of size $M \times K$, the weight matrix of size $K \times N$ is read from GPU global memory to compute the $M \times N$ output matrix (Figure~\ref{fig:gemm}).
Mixture-of-experts (MoE) models route tokens to each expert's weight matrix, but each expert performs the same matmul, so the same memory access pattern applies.

\label{sec:bg_tiled}
\parhead{Tiled Matrix Multiplication.}
GPU matmul implementations partition the matrices into \emph{compute tiles} as shown in Figure~\ref{fig:gemm}, each assigned to a cooperative thread array (CTA) for parallel execution~\cite{nvidia2026cublas, nvidia2026cutlass, tillet2019triton}.
At each iteration over the $K$ dimension, a tile of the weight matrix is loaded from GPU global memory into SRAM to compute a partial matmul.
Compute tile dimensions typically range from 32 to 512 elements.

\subsection{Data Layout in GPU Memory System}
\label{sec:bg_layout}

\begin{figure}[htb]
    \centering
    \includegraphics[width=1.0\linewidth]{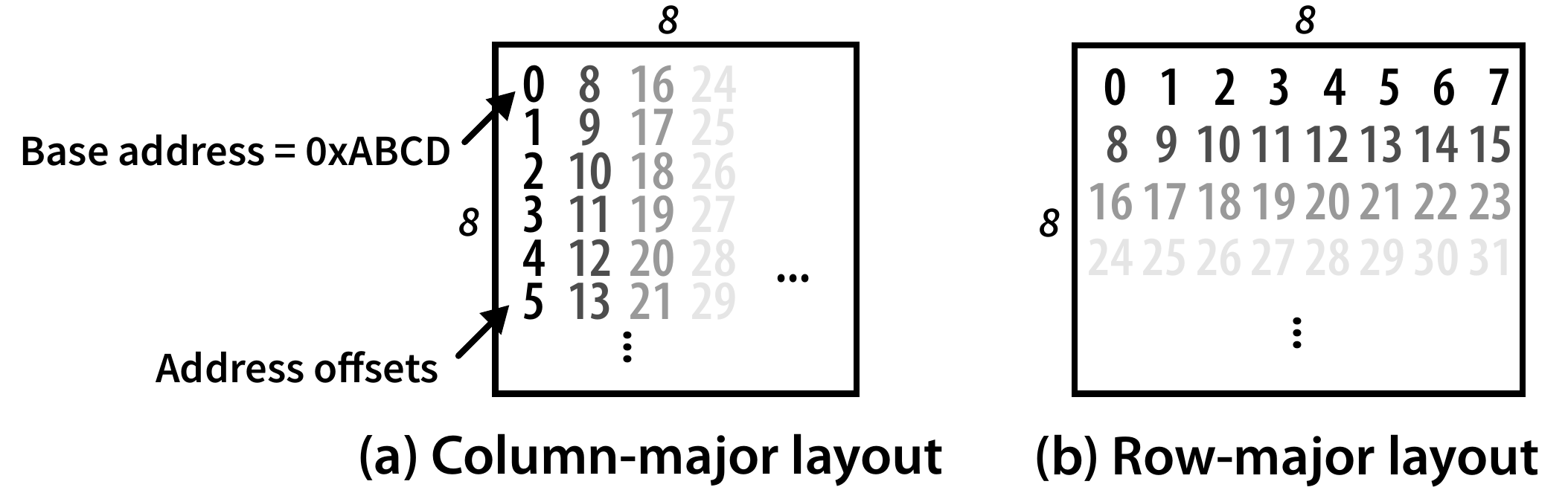}
    \vspace{-0.2in}
    \caption{Address offsets under (a) column-major and (b) row-major layout.}
    \label{fig:colmajor}
\end{figure}

Row-major and column-major are the two standard global memory layouts for matrix multiplication (Figure~\ref{fig:colmajor}), keeping the elements of a row or of a column contiguous in memory.
For LLM weight matrices, column-major, which keeps the K-dimension in Figure~\ref{fig:gemm} contiguous, is the dominant choice~\cite{paszke2017automatic}. 
PyTorch's \texttt{nn.Linear} defaults to column-major, and certain quantization formats on NVIDIA GPUs require it~\cite{nvidia2026ptx, nvidia2026cutlass}.

\subsection{Challenges of Employing LGMS}
\label{sec:challenges}

We show that LGMS causes severe read amplification in tiled matrix multiplication.
Because the weight matrix is stored in a one-dimensional layout (row- or column-major), each 4\,KB access returns a contiguous strip along a single dimension, most of which falls outside the two-dimensional compute tile boundary (Figure~\ref{fig:intro_main}-(b)).
Consider a $128 \times 128$ compute tile with FP8 elements: each row of the tile requires 128\,B, but a single 4\,KB memory access fetches $4096 / 128 = 32\times$ more data than needed. With smaller tile sizes, amplification reaches 64--128$\times$.
This read amplification degrades performance through (1) wasted memory bandwidth and (2) straggler-induced CTA stalls.

\parhead{Bandwidth waste.} 
When the extra data from read amplification is not reused, it is discarded, directly wasting memory bandwidth.
Retaining it on-chip for future reuse is infeasible at the required scale: on an NVIDIA H200 GPU with 132 SMs, a $32\times$ amplification factor over 128\,KB of shared memory per SM produces 528\,MB of amplified data per iteration, far exceeding the 50\,MB L2 cache size.

\begin{figure}[h]
    \centering
    \includegraphics[width=1.0\linewidth]{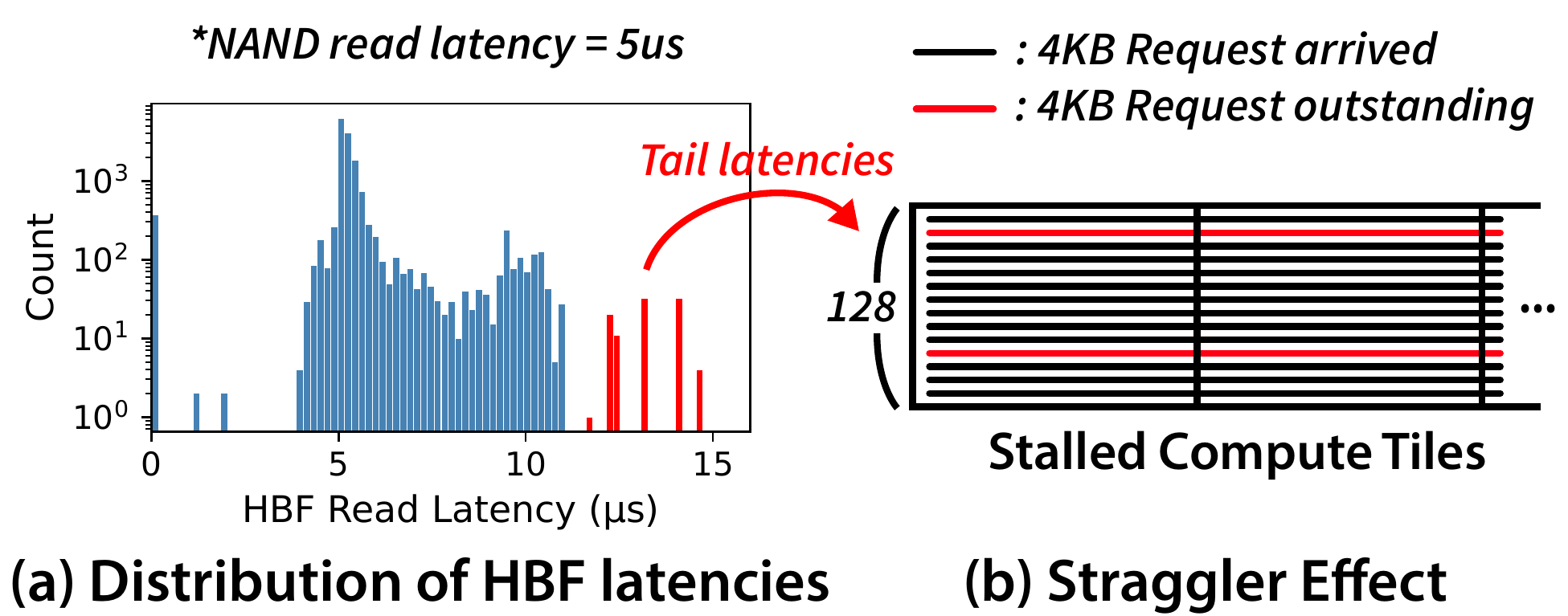}
    \vspace{-0.25in}
    \caption{(a) Log-scale distribution of HBF read latencies showing a long tail. (b) A CTA must wait for all outstanding memory requests to complete before starting to compute; the slowest (straggler) request determines execution time.}
    \label{fig:latency_dist}
\end{figure}

\parhead{Straggler-induced CTA stalls.}
A single compute tile requires multiple 4\,KB memory requests, and the CTA cannot begin computation until all of them complete.
As Figure~\ref{fig:latency_dist} shows, LGMS access latencies exhibit a long tail: the slowest outstanding request determines when each CTA can begin executing.

\begin{figure}[hbt]
    \centering
    \includegraphics[width=0.9\linewidth]{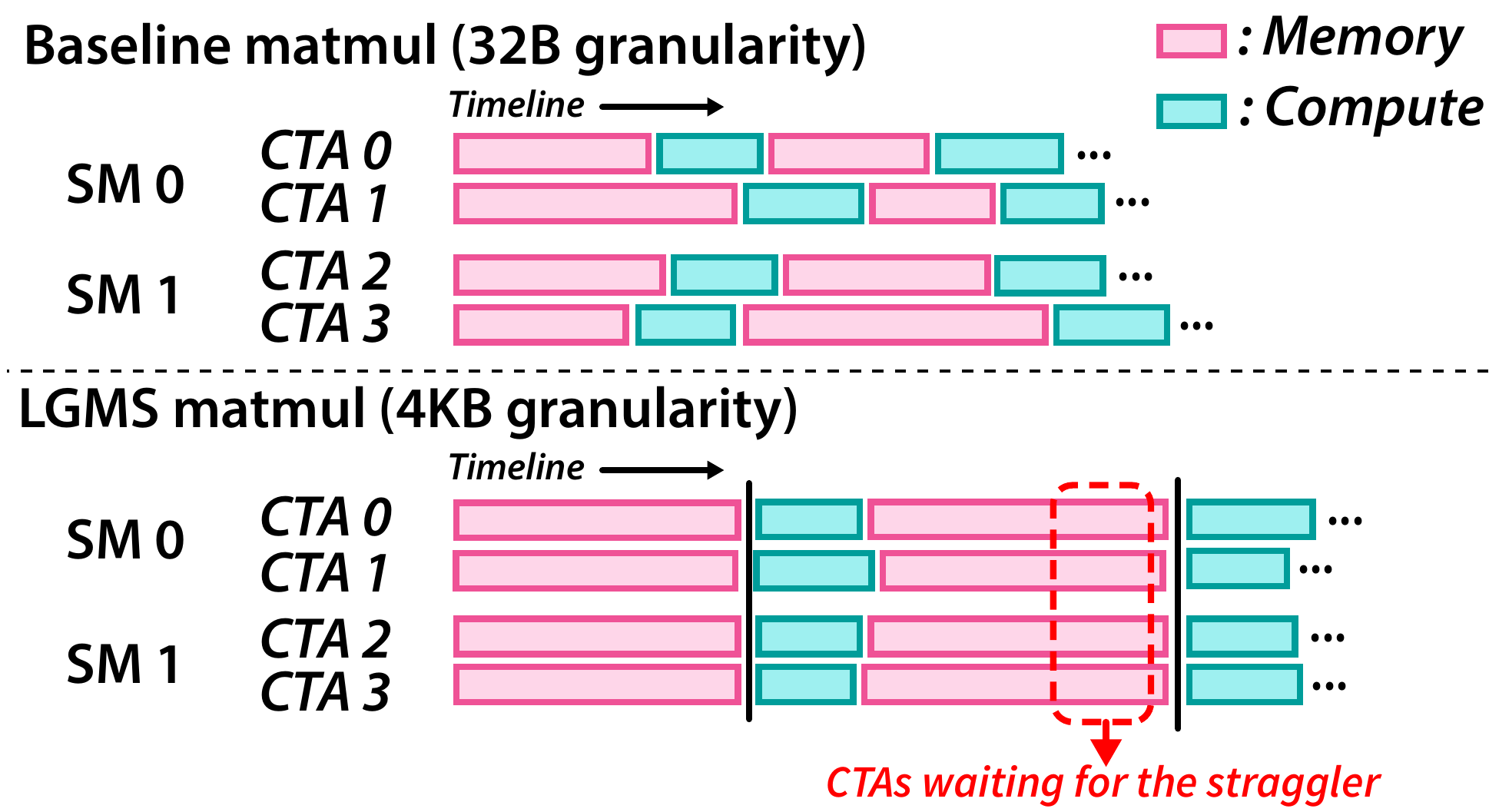}
    \vspace{-0.1in}
    \caption{CTA execution timelines under baseline HBM (top) and LGMS (bottom). CTAs share the same 4\,KB requests, creating implicit synchronization; the straggler request stalls all CTAs that depend on it.}
    \label{fig:timeline}
\end{figure}

This effect compounds across CTAs.
When multiple CTAs share 4\,KB requests due to overlapping tile regions, the straggler request becomes an implicit synchronization point: all dependent CTAs stall until it completes, even though they are otherwise independent.
As shown in Figure~\ref{fig:timeline}, this implicit synchronization serializes CTA execution and stretches the total kernel runtime.

\begin{figure}[t]
    \centering
    \includegraphics[width=1.0\linewidth]{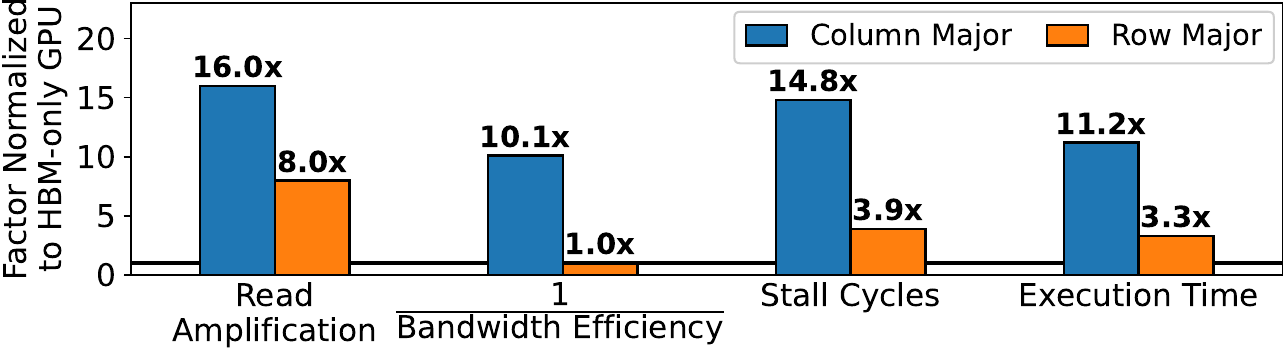}
    \vspace{-0.2in}
    \caption{Performance degradation due to read amplification on LGMS. 
    Metrics measured on Qwen-3 \texttt{fused\_moe} kernel on a hybrid HBM/HBF-stack GPU.
    While memory bandwidth does not get wasted with row-major, straggler effect causes long stalls that underutilize the bandwidth.
    }
    \label{fig:read_amp}
\end{figure}

Figure~\ref{fig:read_amp} quantifies the impact of stragglers on Qwen-3 30B \texttt{fused\_moe} with a $128 \times 256$ FP16 compute tile on a simulated HBF system (setup in Section~\ref{sec:eval_method}).
Column-major layout suffers $10.1\times$ bandwidth waste due to read amplification, which increases stall cycles by $14.8\times$ and degrades execution time by $11.2\times$ relative to an HBM baseline.
Row-major layout does not waste bandwidth because the overfetched row is used by other CTAs, but straggler-induced stalls still degrade execution time by $3.3\times$.

\label{sec:motivation}

\subsection{Goal of the Paper}
Existing works that propose LGMS for GPUs~\cite{ha2026h, hsu2026haven, nam2026rome} focus on scaling memory capacity and bandwidth but do not address read amplification arising from tiled matmul access patterns.
This paper does not advocate for any specific LGMS technology; rather, we identify read amplification as a general consequence of large access granularity and propose a solution that applies to any LGMS.
Our goal is to obtain the capacity and bandwidth benefits of LGMS while eliminating read amplification, with only lightweight changes to GPU software and hardware for practical adoption.
We focus more on HBF as its longer access latency combined with read amplification makes the problem more challenging.

\section{Tile-Major Data Layout}
\label{sec:tilemajor}

\begin{figure*}[t!]
    \centering
    \includegraphics[width=1.0\linewidth]{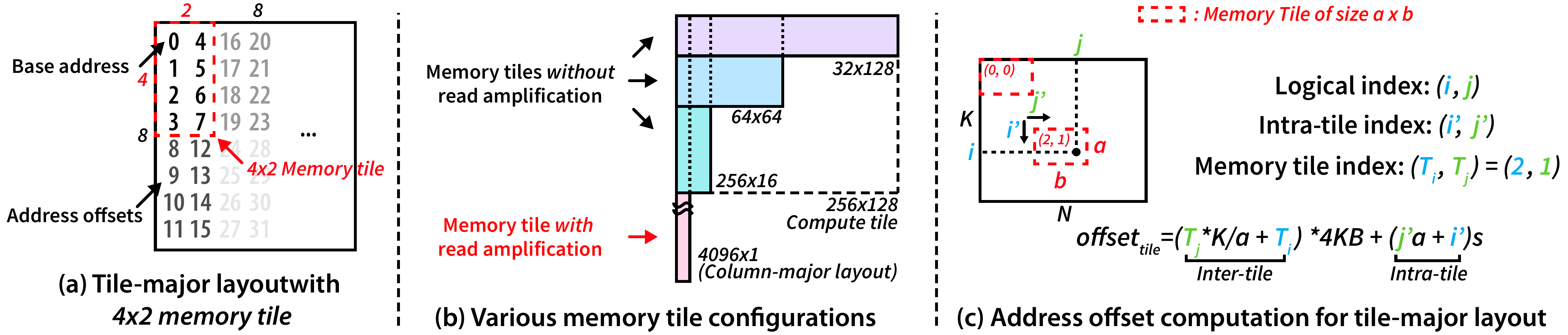}
    \vspace{-0.25in}
    \caption{(a) Address offsets of an $8 \times 8$ matrix under tile-major layout with $4 \times 2$ memory tile. (b) Memory tile shapes for a 4\,KB region relative to a compute tile (dashed). Memory tiles beyond the compute tile ($4096 \times 1$) cause read amplification; shapes that fit within (e.g., $64 \times 64$) eliminate it. (c) Tile-major address translation: element $(i, j)$ is decomposed into an inter-tile index $(T_i, T_j)$ and an intra-tile index $(i', j')$. The tile-major offset is a linear combination of inter-tile index and the intra-tile offset.}
    \label{fig:tilemajor}
\end{figure*}

Read amplification arises because the one-dimensional global memory layout produces 4\,KB strips that do not align with two-dimensional compute tiles.
Therefore we propose to use \emph{tile-major layout}, which reshapes the contiguous unit of global memory from a one-dimensional strip into a two-dimensional rectangle that fits within a compute tile.
We call this rectangular unit the \emph{memory tile} (Figure~\ref{fig:tilemajor}-(a)).

\subsection{Decoupling Memory Tile and Compute Tile}
\label{sec:tilemajor_tiles}

A contiguous 4\,KB region can be shaped as any $a \times b$ rectangle satisfying $a \times b \times s = 4096$, where $s$ is the element size in bytes.
For FP8 ($s = 1$), row-major corresponds to the extreme $(1, 4096)$ and column-major to $(4096, 1)$; both extend far beyond compute tile boundaries, causing read amplification.
By selecting a shape that fits within a compute tile, e.g., $(64, 64)$, every byte in each 4\,KB fetch is consumed by the kernel, eliminating read amplification.

However, the memory layout is fixed at allocation time, while the compute tile shape varies across kernels (e.g., $128 \times 128$ vs.\ $256 \times 64$), so a memory tile matched to one kernel would not work for others.
Tile-major layout therefore \emph{decouples} the memory tile from the compute tile (Figure~\ref{fig:tilemajor}-(b)).
The memory tile is chosen small enough to divide evenly into many compute tile sizes.
For example, a $64 \times 64$ FP8 memory tile fits within compute tiles of $64 \times 64$, $128 \times 128$, $256 \times 256$, and so on.

\subsection{Address Translation}
\label{sec:tilemajor_addr}

We use column-major as the baseline in our explanation; the same approach applies to row-major, with rows and columns swapped. In column-major layout, element $(i, j)$ of a $K \times N$ matrix has byte offset $(j \cdot K + i) \cdot s$ from the matrix base address, where $s$ is the element size in bytes.
As illustrated in Figure~\ref{fig:tilemajor}-(c), for a memory tile $a \times b$ (where $a \cdot b \cdot s = 4096$), we decompose $(i, j)$ into \emph{inter-tile index} $(T_i, T_j)$ indicating which memory tile an element belongs to, and \emph{intra-tile index} $(i', j')$ indicating its position \emph{within that tile}:
\begin{equation}
    T_i = \lfloor i/a \rfloor, i' = i \bmod a,
    T_j = \lfloor j/b \rfloor, j' = j \bmod b.
    \label{eq:tile_indices}
\end{equation}

The tile-major byte offset from the matrix base address is:
\begin{equation}
    \text{offset}_\text{tile} = \underbrace{(T_j \cdot \tfrac{K}{a} + T_i)}_{\text{linear memory tile index}} \cdot \underbrace{a \cdot b \cdot s}_{4\text{\,KB}} \;+\; \underbrace{(j' \cdot a + i') \cdot s}_{\text{intra-tile col-major offset}}
    \label{eq:tilemajor_offset}
\end{equation}
The first term locates the memory tile (tiles are ordered column-major), and the second locates the element inside it. Because $a$ and $b$ are always powers of two ($a \cdot b \cdot s = 4096 =2^{12}$), decomposing $(i, j)$ into inter/intra tile coordinates requires only bit shifts and masks (e.g., $T_i = i \gg \log_2 a$, $i' = i \;\&\; (a{-}1)$).

\subsection{Memory Alignment Requirements}
\label{sec:design_alloc}
We assume the memory allocator aligns the matrix base address to the LGMS granularity and that virtual-to-physical address translation is also aligned to the LGMS granularity.
GPUs use huge pages on the order of MB to reduce TLB pressure, and even the largest memory granularity of tens of KB will fit.
We also assume that the weight matrix dimensions are evenly divisible by the corresponding memory tile dimensions.
The dimensions of recent LLMs (Qwen~\cite{yang2025qwen3}, DeepSeek~\cite{liu2024deepseek}, Llama~\cite{grattafiori2024llama}, Minimax~\cite{li2025minimax}, GPT-OSS~\cite{agarwal2025gpt}) are all multiples of 256.
If the alignment is not satisfied, weights can be padded.

\subsection{Challenges of Adopting Tile-Major Layout}
\label{sec:tilelenshw_challenges}
Unlike row/column major, tile-major is not a single layout but a family of layouts parameterized by the memory tile dimensions $(a, b)$.
Figure~\ref{fig:tilemajor}-(b) shows three valid memory tile shapes with the same total size.
Moreover, large memory granularity makes kernel performance sensitive to the memory tile shape, making it a tunable parameter.
Since the correctness of a kernel depends on the global memory layout, a separate kernel is needed for each memory tile shape.
\rev{This imposes a significant programming burden, since every kernel that reads a tile-major tensor, including all variants an autotuner selects across batch sizes, must be rewritten.
For NVIDIA Hopper (SM90) alone, cuBLASLt ships 2,920 GEMM kernels as closed binaries, and the CUTLASS 3.x generator emits 20,864 GEMM kernel configurations.}
Transparent support in GPU software and hardware would eliminate this effort and let legacy kernels use tile-major without changes, easing the transition to LGMS-based GPUs.

\parhead{Why conventional remapping methods fall short.}
Remapping addresses where it conventionally happens, e.g. in the page table, does not suffice.
A page table can relocate a 4\,KB page but cannot change which elements fall within it, so each page still holds 1-D strips.
Hardware could instead correct each 1-D address into a tile-major address, which requires recovering the logical indices from the linear offset.
But such correction places non-constant division and modulo operations on the critical path of every memory access, since LLM shapes are typically not powers of two.
The correction is not needed if translation to tile-major happens during address generation from logical indices.

\section{\papername: Transparent Tile-Major Layout}
\label{sec:design}

We propose \emph{\papername}, lightweight extensions to GPU software and hardware that transparently enable tile-major layout in GPU global memory.
Our observation is that a dedicated hardware that computes memory addresses from \emph{logical tensor coordinates} serves as a natural transition point for tile-major: most code remains agnostic to the global memory layout, and the remapping avoids the division and modulo.
Such hardware already exists across GPU architectures, including the NVIDIA Tensor Memory Accelerator (TMA)~\cite{nvidia2024h100}, AMD shim DMA~\cite{amd2026am020}, and Intel 2D block loads~\cite{intel2025simd}.
We demonstrate our approach on the TMA, widely used to load matmul weights on recent NVIDIA GPUs, but the same approach applies to the other accelerators.

We present three adoption paths: \emph{\papername-SW} enables DSL-based kernels to use tile-major by changing only the layout descriptor (Section~\ref{sec:design_kernel}); \emph{\papername-HW} extends the TMA hardware so that non-DSL kernels, including low-level binaries, gain transparent support (Section~\ref{sec:design_tma}); and kernels that use neither TMA nor the DSL can adopt tile-major layout by modifying their address computation, either in source code or through compiler-based binary instrumentation (Section~\ref{sec:design_manual}). 
\rev{
All three paths apply to writes as well as reads, since stores compute their global-memory addresses with the same layout descriptor, TMA logic, or address computation.}

\subsection{Tensor Memory Accelerator (TMA)}
\label{sec:bg_tma}
As GPU compute throughput and memory bandwidth grew, computing memory addresses from logical indices began to limit performance~\cite{luo2025dissecting}.
The TMA, introduced in NVIDIA Hopper~\cite{nvidia2024h100}, is dedicated hardware that computes addresses and moves data asynchronously with the GPU core, so data transfer overlaps with compute.
High-performance kernels rely on this overlap and use TMA widely~\cite{spector2024thunderkittens}.

\begin{figure}[t]
    \centering
    \includegraphics[width=1.0\linewidth]{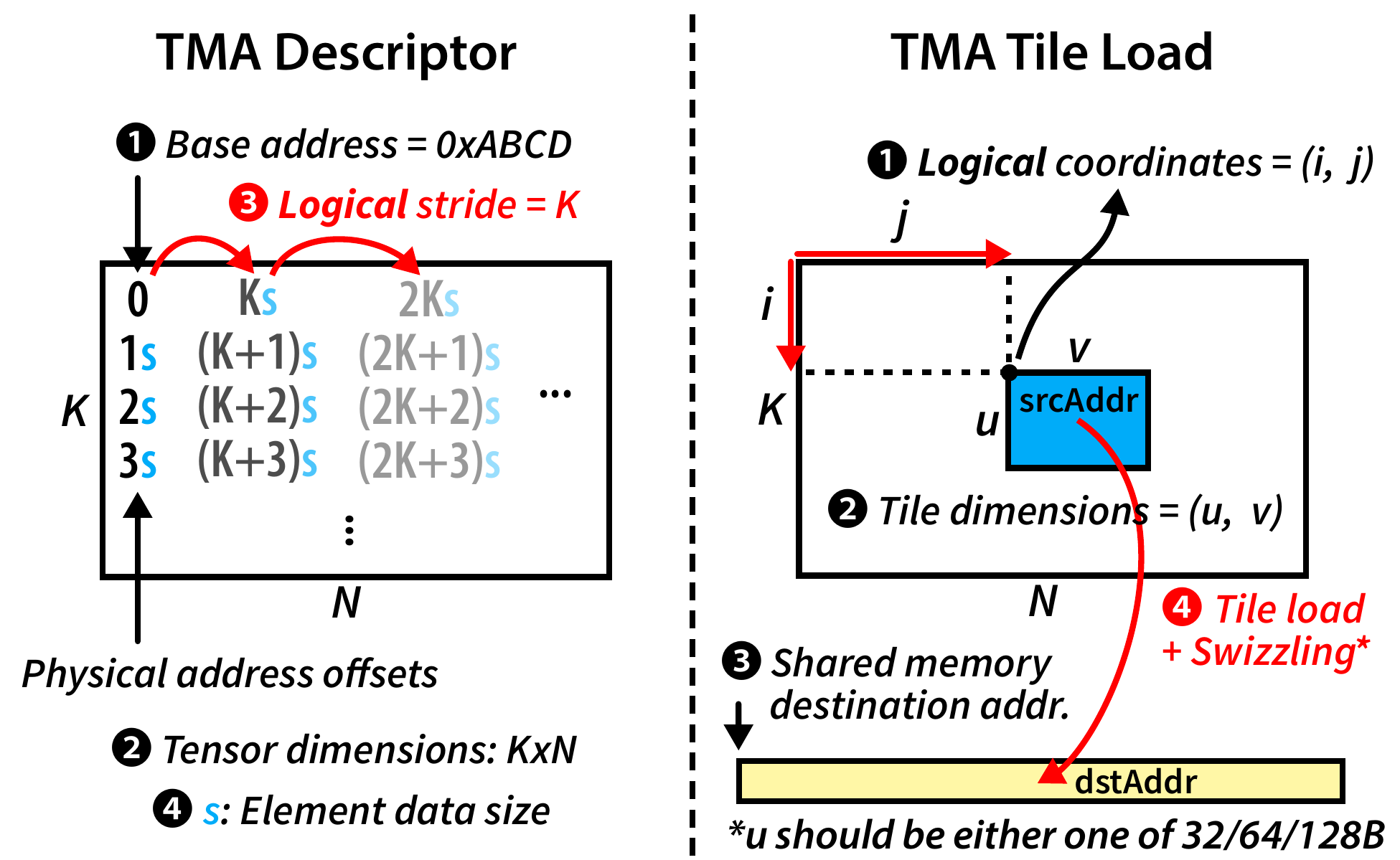}
    \vspace{-0.2in}
    \caption{TMA descriptor and TMA load operation.}
    \label{fig:tma_explanation}
\end{figure}

Figure~\ref{fig:tma_explanation} illustrates the input and operation of the TMA.
Before kernel launch, the host creates a \emph{TMA descriptor} that specifies the global tensor's (1)~base address, (2)~dimensions (e.g. $(K, N)$), (3)~stride for each dimension (e.g. $(1, K)$), and (4)~data type.
During kernel execution, the TMA load is issued with (1) logical coordinates of the tile base $(i, j)$, (2) tile dimensions $(u, v)$, and (3) shared memory destination address.
The TMA computes global memory addresses for data in the tile, issues memory requests, writes the returned data to shared memory, and signals the core when done.

\parhead{Tensor Strides and Address.}
\emph{Stride} specifies how much the memory address changes when the corresponding coordinate is incremented by one.
In Figure~\ref{fig:tma_explanation}-Left, the stride is $K$ in the horizontal dimension.
Given a set of coordinates ($i_1, i_2, ...$) and corresponding strides ($s_1, s_2, ...$), the memory address is the sum of each coordinate multiplied by its stride ($i_1s_1+i_2s_2+...$), plus the base address.
TMA assumes the first dimension is contiguous (stride=1), so for an $N$-dimensional tensor, $N{-}1$ stride values are specified in the descriptor.

\subsection{\papername-SW: Tile-Major via DSL Extension}
\label{sec:design_kernel}
Kernel libraries such as CUTLASS~\cite{nvidia2026cutlass} build on layout abstractions such as CuTe, which decouple compute logic from memory layout and allow kernels to support different layouts by changing only the layout descriptor.
In CuTe, tile-major can be expressed by viewing a $K \times N$ matrix as a 4-D tensor with the memory tile $(a, b)$ at the front:

\vspace{-0.2in}
\begin{align}
\text{shape} &= (a,\; b,\; K/a,\; N/b), \nonumber \\
\text{stride} &= (1,\; a,\; ab,\; Kb),
\label{eq:4d_layout}
\end{align}

where $a \cdot b \cdot s = 4096$ for a 4\,KB granularity memory.
The two additional dimensions separate intra-tile and inter-tile indices.
The stride of the third dimension is $ab = 4096/s$ elements, which corresponds to exactly one 4\,KB memory tile, so adjacent tiles are contiguous in memory and aligned to the memory granularity.

\rev{However, existing CUTLASS kernels cannot directly adopt tile-major, because their code assumes row- or column-major global memory layouts even though CuTe can express arbitrary layouts.
For example, kernels would treat the matrix shape as a list of integers, e.g., divide on shapes to compute the number of tiles, rather than an abstract layout. Such code will fail to run if we pass multi-dimensional views like Equation~(\ref{eq:4d_layout}).

\papername-SW solves this by passing the memory tile information in the stride instead of the shape.
Unlike the matrix shape which is used in many parts of the kernel, its stride is only used to build the abstract layout with CuTe API's \texttt{make\_layout(shape, stride)}.
We therefore pass tile-major stride in Equation~(\ref{eq:4d_layout}) in the kernel argument, and extend \texttt{make\_layout()} to infer and build the matching tile-major shape from it.
Then, the resulting layout becomes tile-major, and the rest of the kernel runs on it unchanged.

At runtime, \papername-SW and CuTe convert 2-D coordinates $(i,j)$ to 4-D coordinates $(i', j', T_i, T_j)$ via Equation~(\ref{eq:tile_indices}), and apply the conversion to the TMA load call.
CUTLASS kernels thus adopt tile-major without any change to their code, and only two parts of CuTe change: layout construction and TMA call generation.
We verified on an H200 GPU that unmodified CUTLASS kernels (dense TMA and \texttt{cp.async} GEMMs, grouped/MoE GEMMs including TMA multicast, and \texttt{FP8} blockwise-scaled GEMMs) with tile-major layout on CuTe produce results identical to the row-/column-major baselines.}

\subsection{\papername-HW: TMA Extension for Tile-Major Layout}
\label{sec:design_tma}
In this section, we propose \emph{\papername-HW}, an extension to address computation hardware like TMA to provide transparent tile-major support for non-DSL kernels.
With \papername-HW all kernels that utilize TMA can be switched to use tile-major by only providing a different TMA descriptor at runtime.
Our insight is that because TMA operates on \emph{logical tensor indices} rather than memory addresses, the index-to-address mapping is fully contained within the TMA. Thus it is a natural place to apply tile-major remapping.
With our extension the TMA performs tile-major address remapping internally, so a single kernel binary supports different memory tile configurations without recompilation.
We only modify the TMA's source address (\texttt{srcAddr}) computation logic.

\begin{figure}[h]
    \centering
    \vspace{-0.1in}
    \includegraphics[width=1.0\linewidth]{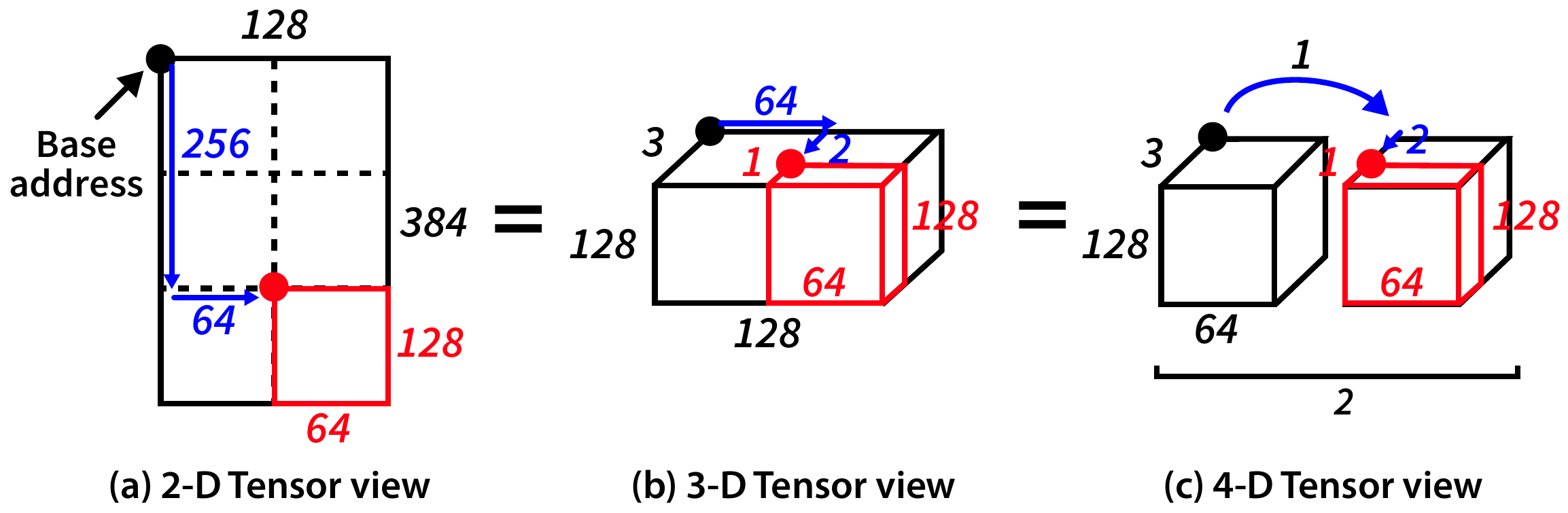}
    \vspace{-0.25in}
    \caption{A single 384$\times$128 column-major matrix and the 128$\times$64 tile at logical coordinates (256, 64), viewed as equivalent (a) 2-D, (b) 3-D, and (c) 4-D TMA tensor descriptors.}
    \label{fig:view}
\end{figure}

\parhead{Challenges of Supporting Tile-Major in TMA.}
Transparent hardware support is challenging because programmers can \emph{view} the same tensor in multiple ways, and the TMA must produce correct tile-major addresses for every view.
Consider an FP8 $384 \times 128$ matrix in column-major shown in Figure~\ref{fig:view}, and the kernel wants to load a $128 \times 64$ tile at logical coordinates $(256, 64)$.
This load can be expressed as form in the following table.
{
\begin{center}
\begin{tabular}{lccc}
\toprule
 & Tensor & Coordinates & Tile \\
\midrule
2D & $(384, 128)$ & $(256, 64)$ & $(128, 64)$ \\
3D & $(128, 128, 3)$ & $(0, 64, 2)$ & $(128, 64, 1)$ \\
4D & $(128, 64, 3, 2)$ & $(0, 0, 2, 1)$ & $(128, 64, 1, 1)$ \\
\bottomrule
\end{tabular}
\end{center}
}

Moreover, except for the contiguous first dimension, the programmer is free to reorder the remaining dimensions, along with their strides, tile sizes, and coordinates.
Kernels routinely use the different views to simplify coordinate computation from the CTA index and iteration count.
A viable hardware solution must handle all these representations with minimal overhead, without a separate circuit for each.

\begin{figure}[t]
    \centering
    \includegraphics[width=0.9\linewidth]{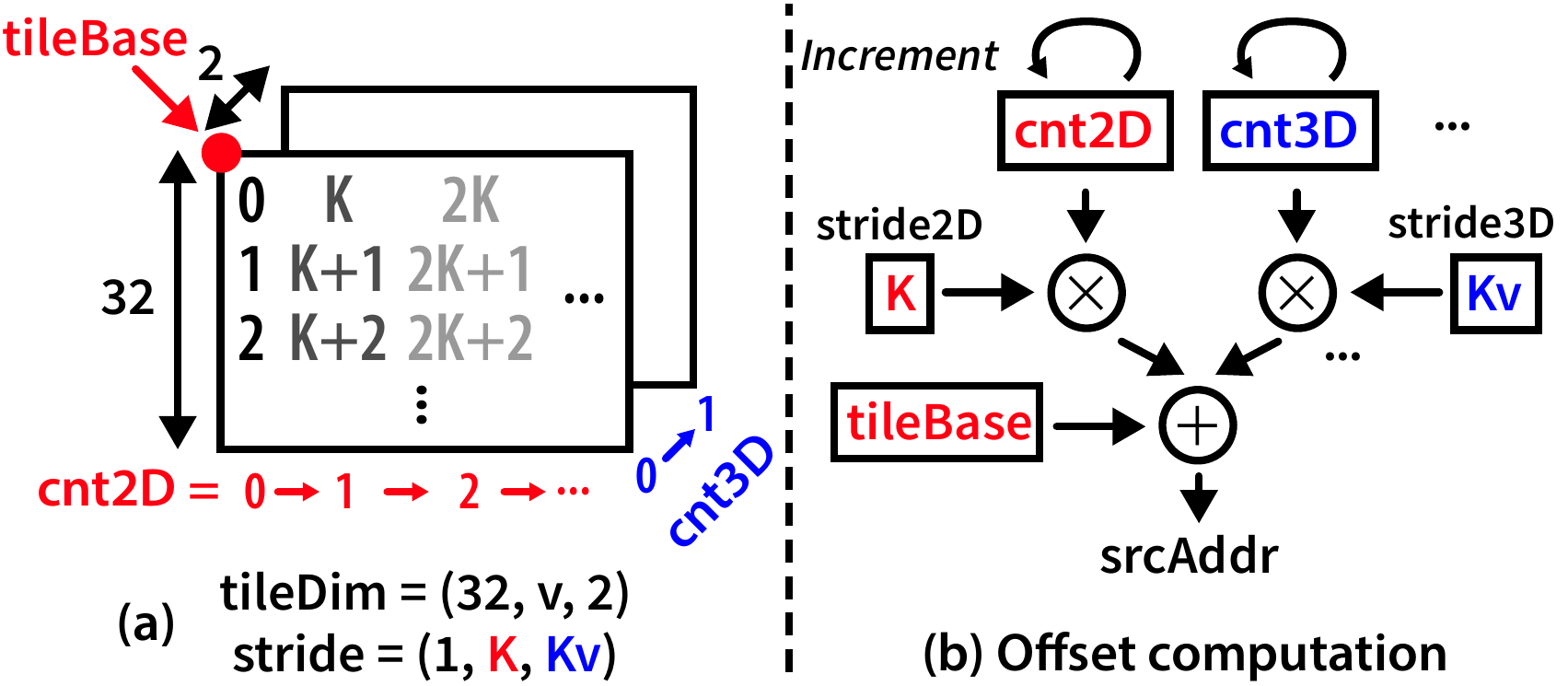}
    \vspace{-0.1in}
    \caption{srcAddr computation: (a) counters sweep the tile dimensions, (b) srcAddr computation through multiplications and additions of counters, strides, and tileBase.}
    \label{fig:tma_counter}
\end{figure}

\begin{figure}[b]
    \centering
    \includegraphics[width=1.0\linewidth]{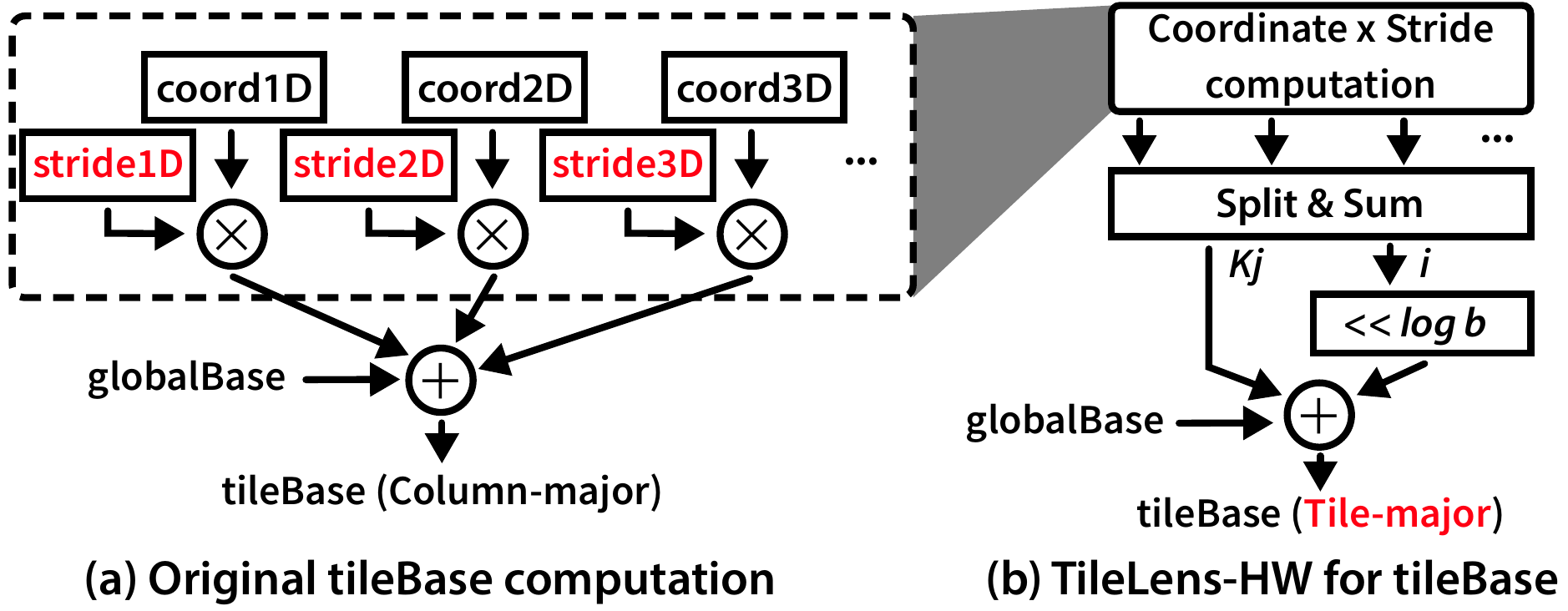}
    \vspace{-0.25in}
    \caption{(a) Computation of the tile base address from coordinates and strides. (b) \papername-HW's extension for the tile-major address computation reuses most existing logic.}
    \label{fig:tma_base}
\end{figure}

\parhead{TMA Microarchitecture.}
While the exact microarchitecture of TMA is not publicly documented, we assume a simple counter-based hardware similar to DMA engines that support multi-dimensional data transfer~\cite{ti2012edma3, zhou2025tensor}.
We model the TMA operation as two phases.
First, TMA computes the tile base address (\texttt{tileBase}) from the coordinates and strides (Figure~\ref{fig:tma_base}-(a)).
Second, a hardware unit with incrementing counters generates source and destination addresses (\texttt{src/dstAddr}) and issues memory requests (Figure~\ref{fig:tma_counter}-(a, b)).
The counters represent the coordinates within the TMA tile, and the source address is a sum of the counters multiplied by corresponding strides, plus \texttt{tileBase}.
The destination address simply increments in the coordinate order.

\parhead{Invariants in TMA calls.}
Two invariants in TMA calls enable a hardware extension with minimal overhead.
First, the contiguous dimension $u$ of a \emph{TMA tile} is a power of two in bytes (except for sub-byte types like FP6~\cite{nvidia2026cutensormap}).
This is required for shared-memory swizzling, which is essential for matmul performance~\cite{spector2024thunderkittens} and permutes bits within a power-of-two width.
Since the memory tile dimensions $a$ and $b$ are also powers of two, one divides the other, guaranteeing alignment in the contiguous dimension.

Second, for all views of a column-major $K \times N$ matrix, each dimension either indexes within a column (stride $< K$) or across columns (stride $\geq K$). This invariant gives hardware a simple rule for splitting arbitrary multidimensional coordinates into row and column indices. We call $K$ the \emph{leading stride}.

\parhead{\papername-HW.}
\papername-HW adds two fields to the TMA descriptor: (1)~the memory tile dimensions $(a, b)$, both powers of two, where $a$ is the contiguous dimension, and (2)~the leading stride.
Using these fields, the TMA engine performs tile-major remapping during its two phases: \texttt{tileBase} address computation and counter-based \texttt{srcAddr} generation within the tile.
We describe the required modifications to each phase under two cases.

\begin{figure}[t]
    \centering
    \includegraphics[width=1.0\linewidth]{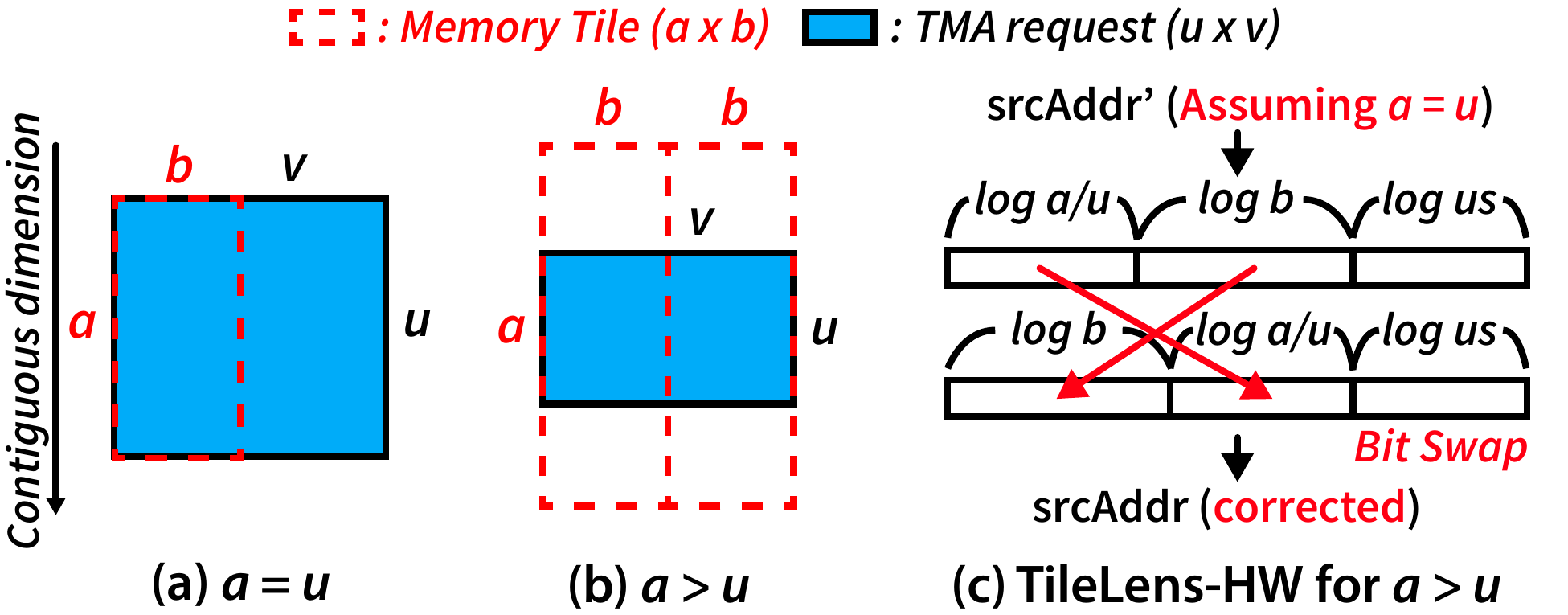}
    \vspace{-0.25in}
    \caption{Two cases of tile-major with memory tile of (a, b) on TMA loading (u, v) tile. (a)~$a=u$. (b)~$a > u$. (c) \papername-HW logic for $a>u$ case.}
    \label{fig:tma_normal}
\end{figure}

\parhead{Case~1: Aligned Tiles.}
In the simplest case, the memory tile and TMA tile have the same contiguous width ($a = u$), and the TMA tile spans an integer number of memory tiles in the non-contiguous dimension (Figure~\ref{fig:tma_normal}-(a)).

To compute \texttt{tileBase}, we use the leading stride to split the multidimensional coordinates into row index $i$ and column index $j$ (Figure~\ref{fig:tilemajor}-(c)): dimensions with a stride smaller than the leading stride contribute to $i$, and the rest contribute to $j$.
In hardware, this is a split and sum unit (Figure~\ref{fig:tma_base}-(b)), where each coordinate$\times$stride product is routed to a different addition tree based on a comparison with the leading stride.
The rest of the tileBase computation hardware is entirely reused.

Because memory tiles evenly divide the global matrix ($K$ and $N$ are multiples of $a$ and $b$, respectively), $i$ and $j$ are always aligned to memory tile boundaries, and Equation~(\ref{eq:tilemajor_offset}) simplifies to:

\begin{equation}
    \texttt{tileBase} = \text{offset}_{\text{tile}}=(K \cdot j \;+\; i \cdot b) \cdot s.
    \label{eq:tilebase_simple}
\end{equation}

The only difference from the column-major offset ($\text{offset}_\text{col} = (K \cdot j + i) \cdot s$) is that $i$ is multiplied by $b$, which reduces to a single variable bit-shift ($\ll \log_2 b$) (Figure~\ref{fig:tma_base}).

With \texttt{tileBase}, the TMA generates \texttt{srcAddr} for each memory request within the TMA tile.
In a standard column-major load, a single counter steps $v$ times with a stride of $K$ to traverse the non-contiguous dimension.
In tile-major, however, columns within a memory tile are only $a$ elements apart, and after $b$ columns, the counter must jump to the next memory tile, which is $b \cdot K$ elements away.

We handle this strided behavior by splitting the dimension with stride $K$ into two \emph{nested} counters:
\begin{equation}
    \text{strides:~} (1,\; \cdots,\; \underbrace{a,\; b \cdot K}_{\text{replaces } K},\; \cdots), \quad \text{tile:~} (u,\; \cdots,\; \underbrace{b,\; v/b}_{\text{replaces } v},\; \cdots)
    \label{eq:counter_reconfig}
\end{equation}
The rest of the TMA operation is identical: we only change the initialization.

\parhead{Case~2: Wide Memory Tile ($a > u$).}
When the memory tile is wider than the TMA tile in the contiguous dimension ($u < a$, but $u$ divides $a$), the engine computes \texttt{tileBase} and counter strides as in Case~1, as if the memory tile size were $u \times b$ and not $a \times b$.
The only correction needed is a bit permutation after srcAddr computation. Because both $u$ and $a$ are powers of two, the correction from the assumed $u \times b$ sub-tile to the actual $a \times b$ memory tile reduces to a bit permutation within the intra-tile offset. For example, if the TMA tile is $32 \times 32$ but the memory tile is $128 \times 32$ (4\,KB), the 12-bit offset changes from $(\textit{sub-tile}_2 \mid \textit{col}_5 \mid \textit{row}_5)$ to $(\textit{col}_5 \mid \textit{sub-tile}_2 \mid \textit{row}_5)$, where subscripts denote bit widths.

Note that Case~2 does not necessarily lead to read amplification.
For example, if the contiguous dimension of a compute tile is wider than 128\,B, optimized kernels (e.g., cuBLAS~\cite{nvidia2026cublas}) issue multiple TMA loads within the compute tile to apply shared memory swizzling.

\subsection{Power, Area, and Latency Overhead of \papername-HW}
\label{sec:hw_overhead}
The hardware extensions of \papername-HW include the split and sum unit (Figure~\ref{fig:tma_base}-(b)), the stride and tile dimension converter (Equation~(\ref{eq:counter_reconfig})), and bit permutation (Figure~\ref{fig:tma_normal}-(c)). The split and sum unit is the most dominant, requiring up to five (TMA supports up to 5-D) additional 32-bit comparators, adders, and routing logic per SM.

To estimate the overhead of \papername-HW, we implement the added components in RTL, and synthesize using Yosys~\cite{wolf2013yosys} targeting ASAP7~\cite{clark2016asap7}, a predictive PDK for a 7\,nm node.
Per SM, the power consumption is 1.43\,mW and the total cell area is 484.9\,$\mu$m$^2$, which become 189\,mW and 0.064\,mm$^2$ for 132 SMs on an H200 GPU. Even without accounting for the more advanced 4\,nm process used in H200, the power and area overheads are 0.03\% and 0.008\%, respectively.

\papername-HW does not affect the latency between memory requests within a single TMA call, but adds 4 cycles at 2.0GHz to the initial \texttt{tileBase} computation time. 

\subsection{Adoption for Non-TMA/Non-DSL Kernels}
\label{sec:design_manual}

Kernels that use neither TMA nor a DSL can adopt tile-major layout by modifying their address computation in kernel code.
\rev{The column-major offset $(j \cdot K + i) \cdot s$ used for memory access needs to be replaced with $\text{offset}_\text{tile}$ in Equation~(\ref{eq:tilemajor_offset}).
Since the offset is computed from the logical indices $(i, j)$, it requires only bit shifts, masks, and a multiplication, adding minimal overhead per access.
When measured on an H200 GPU, the overhead is less than 1\%.
However, unlike \papername-SW or \papername-HW, the change must be applied separately to each kernel, for each different memory tile shape.
}

\rev{Binary instrumentation is a fallback for binaries \textit{without} source code or TMA (e.g., cuDNN, cuBLAS).
It injects remapping instructions before each global memory access, but only sees the linear offset, so it must divide each offset by $K$ to recover $(i, j)$.
On H200, this takes about 30 cycles for a power-of-two $K$, but 150--170 cycles of Barrett reduction~\cite{barrett1986implementing} otherwise.
With four serial \texttt{LDGSTS} per tile, the latter adds about 700 cycles to an L2 hit.

Validated on our simulator, instrumentation is 7.7\% (Qwen) and 7.8\% (Llama) slower than \papername-HW with 5\,$\mu$s HBF on average.
For non-power-of-2 $K$, Llama's down projection ($K$ = 28,672) and Qwen3's w2 ($K$ = 768) slow down by 12\% and 20\%.
This confirms that tile-major remapping is cheap from logical indices but costly as a correction to linear offsets in instructions.}

\subsection{\rev{General Applicability of \papername Across Vendors}}
\label{sec:requirements}
\rev{
\papername-HW relies on only two properties of a tensor address generator, which other vendors' engines provide with at most small changes.
First, the contiguous dimension of a tile must be a power of two, so that the tile and the memory tile divide each other.
Intel 2-D block loads~\cite{intel2025simd} only support power-of-two block widths~\cite{intel2026blockio}.
The AMD Tensor Data Mover~\cite{amd2026tdm} and shim DMA~\cite{amd2026am020} do not restrict the tile width, so the constraint must be enforced in software.
Second, the engine must receive the tensor base address, strides, size, and the logical tile coordinates, so that it can split the coordinates into row and column indices.
The TMA and Intel 2-D block loads accept all four, so \papername-HW applies directly.
The AMD Tensor Data Mover and shim DMA have the first three but receive \texttt{tileBase} address computed in software, so the split-and-sum logic must run in software, or the engine must accept logical tile coordinates.
\papername-SW requires only a DSL with a layout abstraction, and the change it needs is the nested-stride support.
For example, AMD's Composable Kernel~\cite{amd2026composable} describes each GEMM operand with a tensor descriptor of sizes and strides, which can express the 4-D view of Equation~(\ref{eq:4d_layout}) without changing the kernel body.
}

\section{System Support for HBF-Augmented GPUs}
\label{sec:hbf_system}

Tile-major layout addresses the read amplification common to all LGMS.
Adopting HBF on GPUs, however, raises additional system-level challenges: long read latency, limited write endurance, and coexistence with HBM in a hybrid memory hierarchy.
This section addresses them to provide a complete system baseline for evaluating an HBF-augmented GPU.

\subsection{Allocation and Data Placement}
\label{sec:hbf_alloc}
In a hybrid HBM+HBF system, we place each tensor by its access pattern at allocation time.
Model weights are read-only during inference, so they go to HBF, while activations and the KV cache are both read and written, so they stay on HBM.
Each matmul thus reads its weight from HBF and its input activations from HBM.
\papername does not affect endurance, as tile-major layout changes only the arrangement of the weights written at load time, not the amount of data written.

\rev{
\parhead{Host-to-device transfer.} Converting weights to tile-major is a one-time cost paid at model load. GPU can use HBM as a staging buffer to permute the weights before writing to HBF.
On an H200 GPU, copying all layers from the host takes 1.2\,s for Qwen3 30B (60\,GB), and permuting each layer to tile-major only adds 7\% to this time, which drops to 0.2\% when the permutation is pipelined with the next layer's copy.
}

\subsection{Mixed-Granularity L2 Cache \& MSHR}
\label{sec:hbf_l2}

In an HBM+HBF hybrid system, the L2 cache must handle two different access granularities: 32\,B for HBM and 4\,KB for HBF.
When an HBF read request returns a full 4\,KB block, we always insert the full 4\,KB block into L2.
With tile-major, the entire 4\,KB block is useful, and the 4\,KB insertion improves hit rates for subsequent accesses within the same tile.
Without tile-major, this policy pollutes the L2.
Discarding the excess data instead, however, would waste the fetched bandwidth.
\rev{The L2 is statically way-partitioned, with 14 of its 16 ways for HBF data, so 4\,KB insertions cannot evict HBM data. Misses are routed by address: reads to the weight ranges go to HBF, and all other accesses go to HBM.}

The MSHR structure is also changed. The original 128\,B MSHR entries are quickly filled due to the $\mu$s-scale HBF latencies.
We partition the MSHRs between HBM and HBF and track HBF requests at 4\,KB granularity, increasing the trackable HBF pages by $32\times$.
\rev{Each HBF entry covers a 4\,KB block, and every miss to a line of an in-flight block merges into its entry. 
When a warp misses and its MSHR partition has no free entry, the warp stalls until one frees.
Prefetches allocate no MSHRs; the memory controller tracks in-flight prefetched blocks and drops duplicates.
Each HBF channel issues demand reads before prefetches, and a demand to a block with a pending prefetch merges into it and raises it to demand priority.
When a block returns, all 32 lines fill the L2 and release the waiting requests.
Only one 4\,KB data buffer per HBF channel is needed rather than one for every MSHR entry, as the HBF device assembles each 4\,KB block and sends to the GPU as a whole.

The hardware cost is small: an HBF MSHR entry stores the same information as an MSHR entry for HBM, only for a larger block, and partitioning the L2 needs one bit mask per L2 slice to mark the HBF ways. The per-channel data buffer totals 768\,KB for 192 pseudo-channels, and we add 4 bits to the L2 and MSHR tag for the larger address space of HBF.
We estimate the cost of additional SRAM with CACTI~\cite{balasubramonian2017cacti} on a 22\,nm node as 5.7\,W peak power and 1.0\,mm$^2$ area. H200 GPU uses a 4\,nm node, but even without scaling, the power and area overhead is 0.81\% and 0.12\%, respectively.
}

\subsection{Adaptive Degree Prefetcher for HBF}
\label{sec:design_prefetch}

Even with tile-major layout, HBF's read latency is 10--100$\times$ longer than HBM's.
To hide it, we add an adaptive degree prefetcher to the memory controller that keeps enough requests in flight to saturate the HBF bus.
In a tiled matmul kernel, each CTA iterates along the K dimension in fixed-size steps, so the stride is the compute tile size along K.
For each demand access, the prefetcher speculatively issues $d$ additional read requests at stride offsets ahead of the current position.
The goal is to choose $d$ so that the total number of in-flight requests saturates the HBF bus throughout kernel execution.

At any point during execution, the GPU issues one wave of demand requests per K-iteration, accessing $p_{\text{wave}} = N_{\text{SM}} C_{\text{SM}} p_{\text{tile}}$ requests, where $N_{\text{SM}}$ is the number of SMs, $C_{\text{SM}}$ is the number of CTAs per SM, and $p_{\text{tile}}$ is the number of requests per compute tile.

To saturate a bus with bandwidth $B$ over a NAND read latency of $L$ cycles, we need $p_{\text{wave}} (d+1) \geq BL$, giving:

\begin{equation}
    d = \min\left(\left\lfloor \frac{BL}{p_{\text{wave}}} \right\rfloor, \frac{K}{\texttt{TILE\_K}}\right)
    \label{eq:prefetch_degree}
\end{equation}

The degree is capped at $K / \texttt{TILE\_K}$, the number of K-iterations per tensor, to prevent prefetching beyond the tensor boundary.
In practice, we multiply the first term by $2\times$ to compensate for NAND plane collisions, which empirically prevent all in-flight requests from being serviced simultaneously.

The degree adapts dynamically as CTAs retire.
As CTAs complete execution, $p_{\text{wave}}$ decreases and $d$ increases automatically, ensuring that the remaining CTAs issue enough prefetches to keep the bus saturated.
This is critical at the end of a kernel, where a few straggler requests would otherwise underutilize the HBF bus.
Because $d$ depends only on hardware parameters and $p_{\text{wave}}$, it requires no convergence period and reacts immediately when a CTA retires.

This prefetcher is not a general-purpose mechanism: it exploits the fact that tiled GEMM has a known, fixed access stride equal to the compute tile size.
\rev{For HBF-augmented GPUs, only weights are placed in HBF (Section~\ref{sec:hbf_alloc}), so every HBF access can benefit from this prefetcher as it comes from weight GEMMs with uniform access patterns.}
Designing a prefetcher that handles irregular workloads on HBF-augmented GPUs is an interesting direction, but is orthogonal to this work.

\begin{figure}[!b]
    \centering
    \includegraphics[width=0.95\linewidth]{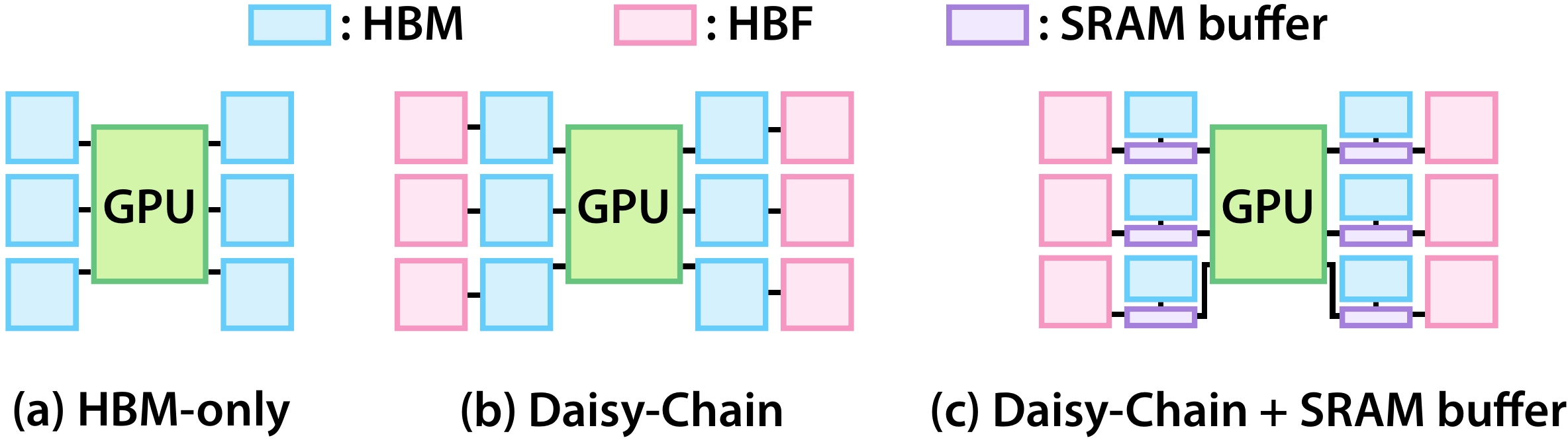}
    \vspace{-0.1in}
    \caption{GPU System configurations for evaluation.}
    \label{fig:system}
\end{figure}

\begin{figure*}[t!]
    \centering
    \includegraphics[width=1.0\linewidth]{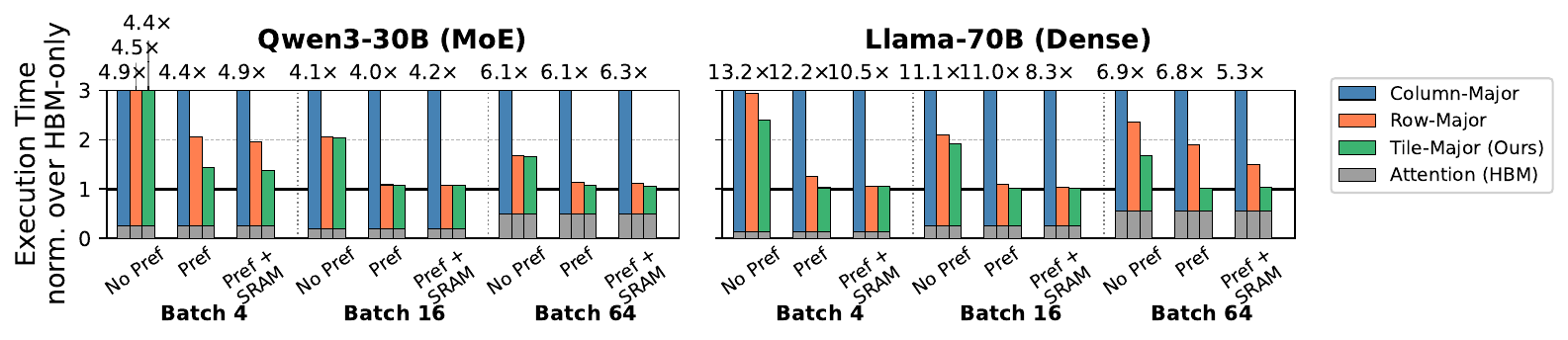}
   \vspace{-0.35in}
    \caption{\rev{Execution time for the decode phase of Qwen and Llama under three global memory layouts. HBF results are shown at 5\,$\mu$s NAND latency with three configurations (No Prefetcher, Prefetcher, Prefetcher+SRAM Buffer). All values are normalized to the HBM-only baseline. Sequence length = 8,192.}}
    \label{fig:main_results}
\end{figure*}

\section{Evaluation}
\label{sec:eval}

\subsection{GPU-HBF System Configurations}
\label{sec:eval_designs}

\parhead{GPU System.}
We evaluate tile-major layout on an H200-class GPU under three memory configurations (Figure~\ref{fig:system}): HBM-only as the baseline, and a daisy-chained HBM+HBF system with and without an SRAM buffer.
In the HBM-only configuration, there are six HBM3e stacks. In the daisy-chained configuration, there are six HBF stacks connected to six HBM stacks, which time-share the data bus.
The daisy-chain design follows H$^3$~\cite{ha2026h}, which also places a 40\,MB SRAM latency-hiding buffer on the HBM base die.
For configurations without the SRAM buffer, the GPU L2 cache is used as a staging area for data from HBF. 
All configurations are derived from NVIDIA H200 and HBM3E specifications~\cite{nvidia2023h200} (Table~\ref{tab:gpu}).

\begin{table}[t]
\centering
\small
\caption{H200-class GPU Parameters used for the evaluation.}
\vspace{-0.1in}
\begin{tabular}{ll}
\Xhline{2\arrayrulewidth}
No. SMs & 132 @ 2.0\,GHz \\
\rev{Warp Scheduler} & \rev{4 schedulers per SM, 1 issue/cycle, RR} \\
\rev{Resident warps per SM} & \rev{up to 64} \\
\rev{CTAs per SM} & \rev{Per-kernel occupancy limit on H200} \\
\rev{Tensor units per SM} & \rev{4} \\
\rev{Tensor core latency~\cite{luo2025dissecting}} & \rev{WGMMA 64 cycles, HMMA 24 cycles} \\
\rev{HMMA throughput} & \rev{721 TFLOPS (pipelined per 6 cycles)} \\
\rev{WGMMA throughput} & \rev{1,081 TFLOPS (FP16 dense)} \\
\rev{ALU latency} & \rev{4} \\
L1 cache per SM & 256\,KB (4-way, 30-cycle hit) \\
L2 cache (total) & 50\,MB (16-way, 270-cycle hit) \\
\Xhline{2\arrayrulewidth}
\end{tabular}
\label{tab:gpu}
\end{table}

When we employ the SRAM buffer in the daisy-chained configuration, our prefetcher targets two tiers: prefetches with a smaller degree (distance 1 through $d$) fill the L2 cache, while far prefetches (distance $d{+}1$ through $2d$) fill the SRAM buffer.

\parhead{Memory Stacks.}
HBM3e has 16 channels per stack with 64\,B granularity (32\,B per pseudo-channel), and the six stacks provide 4.915\,TB/s.
We assume HBF has the same channels and bandwidth, but with 4\,KB granularity.
All data buses run at 3200\,MHz (6.4\,Gbps per pin), with 64 pins per channel.
HBM row buffer hit and miss latencies are 14\,ns and 42\,ns, respectively.
For HBF latency, we sweep the NAND page read latency across 1, 2, 5, 10, and 20\,$\mu$s to cover the range of current and projected flash technologies~\cite{ma2026challenges, sandisk2025hbf, ha2026h, hsu2026haven}.
The total internal bandwidth of HBF is set to $2.5\times$ the bus bandwidth to model concurrent NAND plane access~\cite{koo2017summarizer, dirik2009performance}.
RoMe read latency includes the row miss penalty plus the 4\,KB bus transfer time of 108\,ns.

\subsection{Methodology}
\label{sec:eval_method}
\parhead{Simulator.}
We use Macsim~\cite{kim2012macsim}, a cycle-level GPU simulator, extended with HBM, RoMe, and HBF memory models.
Kernel traces are collected from an NVIDIA H200 GPU using a SASS-level tracer built on NVBit~\cite{villa2019nvbit}.
\rev{
Against Nsight Compute measurements of 364 decode kernels in 24 layer configurations on the H200 GPU, the HBM-only model shows 4.9\% mean and 12.1\% maximum error in decode execution time, with a Pearson correlation of 0.998. 
Per kernel, the MAPE is 10.2\%, 12.8\%, and 17.3\% for FFN, attention, and projection kernels (Pearson $r$ = 0.986, 0.998, and 0.989), which take 94\% of decode execution time. 
Our results are insensitive to compute modeling: sweeping the tensor core latency (0.5--2$\times$) and the ALU latency (2--4$\times$) does not change the relative performance of the layouts; \papername's speedup over baseline changes by 5.1\% in average.}

\parhead{Workloads.} We evaluate the decode (token generation) phase of two representative LLM models with FP16 precision and sequence lengths of 8,192 tokens: a mixture-of-experts (MoE) model Qwen-3 30B-A3B~\cite{yang2025qwen3}, and a dense model Llama-3.1 70B~\cite{grattafiori2024llama}. 
Traces are collected with batch sizes of 4, 16, and 64 for both models, which puts both workloads in a memory-bound regime. All evaluations run a single decode step unless otherwise noted.
\rev{
On an H200, the weight matmuls and attention together take 92--96\% of decode time, while all other kernels take the remaining 4--8\%.
These non-matmul operators (normalization, reductions, silu, etc.) touch only activations and the KV cache in HBM, so tile-major does not affect them with HBF.
Llama's QKV, O, and down projections use a split-K factor of 4 to increase the number of CTAs. Qwen's GEMMs do not use split-K, since their $K$ is small. Section~\ref{sec:eval_kernel_parameter} discusses split-K in more detail.}

\parhead{\papername-HW Performance Modeling.}
\papername-HW adds 4~cycles to TMA initialization (Section~\ref{sec:hw_overhead}), on top of the roughly 170 cycles that a TMA 2D load already spends on initialization~\cite{semianalysis2026microbench}.
Moreover, high-performance kernels overlap TMA loads with computation, so the additional latency is fully hidden in practice.
We therefore do not model any additional \papername-HW latency in our simulation.

\subsection{Kernel Execution Latency per Layout}
\label{sec:eval_main}

We compare the effect of column-major, row-major, and tile-major memory layout on three system configurations: no prefetching (No Pref), adaptive prefetching (Pref), and adaptive prefetching with an SRAM buffer (Pref+SRAM). 
Tile-major layouts consistently outperform linear layouts across LGMS configurations with an adaptive prefetcher (Figure~\ref{fig:main_results}). Column-major layout suffers 4--13$\times$ slowdown across all HBF configurations, even with the SRAM buffer. 
Row-major layout reduces amplification but still incurs a geometric mean (geomean) slowdown of 1.62$\times$ under HBF due to the straggler effect.
Tile-major layout with adaptive prefetching reduces the geomean slowdown to 1.10$\times$; adding an SRAM buffer yields an identical 1.10$\times$, indicating that the prefetcher alone is sufficient to hide HBF latency. The slowdown is concentrated in the smallest batch (Qwen-3 batch 4); every other configuration lands within 8\% of the HBM-only baseline.
Without prefetching, tile-major is no faster than row-major on Qwen, as read amplification in the row-major layout serves as implicit prefetching.
Tile-major eliminates this amplification, so every CTA issues its own demand requests to HBF, unless the prefetcher issues requests ahead of demand to hide HBF latency.

\rev{
\parhead{Effect of batch size.} Tile-major remains the fastest layout at every batch size, whereas row-major depends on how kernels are tuned for each batch size.
In decode, \texttt{TILE\_M} equals the batch size $M$, so a GEMM's CTA count depends on \texttt{TILE\_N} and the split-K factor rather than on $M$.
Most GEMMs keep the same \texttt{TILE\_N} across batch sizes, but some are retuned.
For example, for Llama gate/up GEMMs, the autotuner at batch 64 selects a smaller \texttt{TILE\_N} than at batch 16, so each 4\,KB page is shared by 16 CTAs instead of 8, and each straggler page stalls twice as many CTAs (Section~\ref{sec:eval_straggler}).
These kernels account for more than 90\% of the row-major shortfall at batch size 64 (1.84M of 3.47M cycles, versus 0.41M under tile-major).
Tile-major avoids this problem, since each page belongs to a single CTA regardless of per-batch tuning.

\parhead{Attention kernel's share.} A larger attention share narrows the gap between layouts, since attention reads the KV cache from HBM regardless of the weight layout.
For Llama, weight reads are fixed while KV reads grow with batch size, so the attention share rises from batch 4 to 64.
For Qwen, FFN weight reads also grow from batch 4 to 16 as more experts become active, so the share stays similar; by batch 16 most of the 128 experts are active, so only attention grows afterward.
}

\begin{figure}[t]
    \centering
    \includegraphics[width=1.0\linewidth]{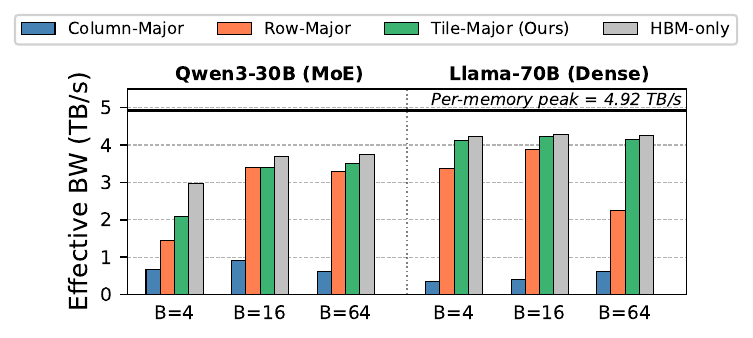}
    \vspace{-0.35in}
    \caption{Effective memory bandwidth utilization at 5\,$\mu$s NAND latency. Peak bus bandwidth for HBM is also shown.}
    \label{fig:bw_results}
\end{figure}

\subsection{Bandwidth Utilization}
\label{sec:eval_bw}

\begin{figure}[t]
    \centering
    \includegraphics[width=0.7\linewidth]{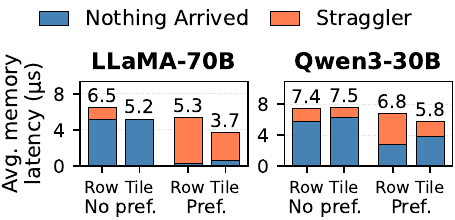}
    \vspace{-0.15in}
    \caption{\rev{Average memory access latency breakdown for the Llama and Qwen's up/gate kernels on 5\,$\mu$s HBF. Tile-major layout reduces the average access latency.}}
    \label{fig:sync_results}
\end{figure}

Figure~\ref{fig:bw_results} shows the effective memory bandwidth throughout a single decode step with a sequence length of 8,192 tokens.
Column-major layout saturates the memory bus with amplified reads, but most fetched data falls outside the compute tile and is evicted from L2 before reuse, reducing effective bandwidth to 7.1--18.6\% of peak (a 5.4--14.0$\times$ waste).
With row-major, the fetched data is reused by other CTAs before eviction, and effective bandwidth is higher than column-major.
However, straggler-induced stalls prevent CTAs from issuing enough concurrent requests to saturate the bus, limiting effective utilization to 29.3--78.9\% on HBF.
Tile-major layout achieves 42.3--86.1\% bandwidth utilization across all HBF configurations, which is the closest to the HBM baseline.

\subsection{Straggler-Induced CTA Stalls}
\label{sec:eval_straggler}

To quantify the straggler effect, we measure the latency of every HBF page a CTA requests within one k-iteration of the gate/up projection, averaged over the kernel's k-iterations. Figure~\ref{fig:sync_results} breaks it into (1) the wait until the first page arrives, dominated by the HBF read latency, and (2) the additional wait until the last page arrives (straggler wait), for \rev{Qwen 30B and Llama 70B} at batch 64 on 5\,$\mu$s HBF.

Without prefetching, both layouts pay the same 5.1\,$\mu$s first-page wait, but row-major adds a 1.32\,$\mu$s straggler wait on Llama, since a CTA needs 64 pages and cannot proceed until the slowest arrives.
Under tile-major, a CTA needs only 2 pages per k-iteration, leaving a near-zero straggler wait (0.06\,$\mu$s).
With prefetching, the straggler wait grows for both layouts as some pages hit L2 while the rest pay the full HBF latency, but tile-major's average latency is still 1.4$\times$ shorter.

\begin{figure}[t]
    \centering
    \includegraphics[width=0.80\linewidth]{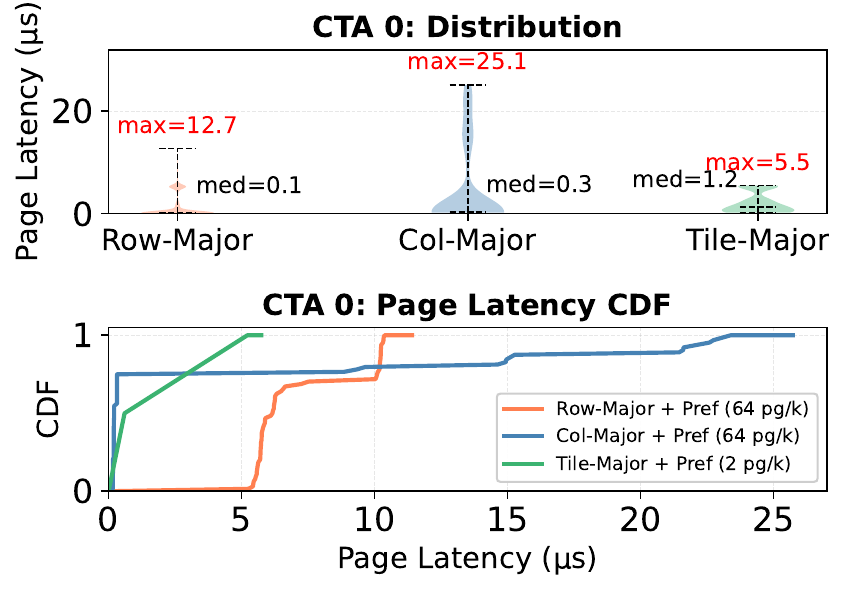}
    \vspace{-0.15in}
    \caption{Per-request latency distribution (top) and CDF (bottom) for a single CTA of Llama up/gate kernel on 5\,$\mu$s HBF with prefetching. Read amplification increases the number of outstanding requests and stragglers.}
    \label{fig:straggler_results}
\end{figure}

Figure~\ref{fig:straggler_results} (top) shows the per-request page latency. Row- and column-major have lower medians than tile-major because other CTAs share their amplified reads through the L2 cache. However, their CTAs need 64 pages per k-iteration where tile-major needs 2, so L2 misses generate more HBF traffic and plane collisions, and the queueing delay stretches the tail.

In the CDF (bottom), tile-major's 2 pages both resolve by 5.2\,$\mu$s, whereas row-major's 64 arrivals spread from 5.3 to 10.4\,$\mu$s and column-major's out to 23.4\,$\mu$s. Since a CTA waits for its slowest page, the tail sets the performance: row-major runs 1.89$\times$ and column-major 6.84$\times$ slower than HBM-only, while tile-major stays within 3\%.

\begin{figure}[t]
    \centering
    \includegraphics[width=1.0\linewidth]{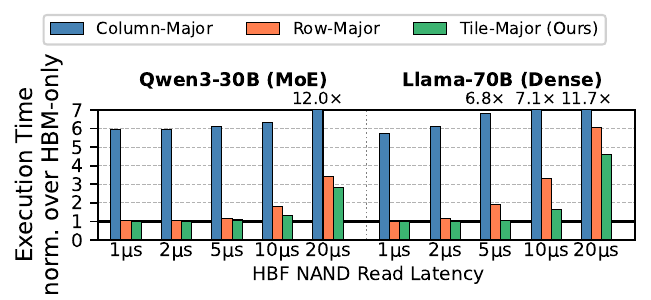}
    \vspace{-0.3in}
    \caption{Execution time sensitivity to HBF NAND read latency (1--20\,$\mu$s). Tile-major with prefetching maintains near-HBM performance up to 10\,$\mu$s.}
    \label{fig:latency_results}
\end{figure}

\subsection{Sensitivity: HBF Latency}
\label{sec:eval_latency}

Figure~\ref{fig:latency_results} shows decode-step execution time as we sweep HBF page read latency from 1 to 20\,$\mu$s at a batch size of 64. Column-major suffers a 5.8--12$\times$ slowdown across all latencies, as read amplification dominates regardless of HBF speed. At 1\,$\mu$s row- and tile-major are indistinguishable since the latency is short enough to tolerate the additional outstanding requests. However, beyond 5\,$\mu$s row-major degrades sharply on Llama, as straggler delays leave CTAs unable to sustain enough parallelism to hide the latency. Tile-major with prefetching stays within 10\% of the HBM baseline through 5\,$\mu$s, but reaches 2.8--4.6$\times$ at 20\,$\mu$s, indicating the prefetcher's coverage limit.

\begin{figure}[t]
    \centering
    \includegraphics[width=0.8\linewidth]{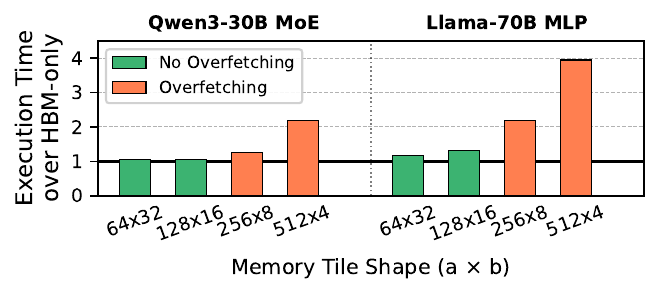}
    \vspace{-0.15in}
    \caption{Kernel execution time for tile-major as memory tile shape varies for a $128 \times 256$ compute tile. Wider tiles fit within the tile boundary; narrower tiles extend beyond it, increasing read amplification and straggler delays.}
    \label{fig:sensitivity_tile_shape}
\end{figure}

\subsection{Sensitivity: Memory Tile Shape and Access Granularity}
\label{sec:eval_shape}

Figure~\ref{fig:sensitivity_tile_shape} sweeps over the memory tile shape for a fixed $128 \times 256$ FP16 compute tile on 5\,$\mu$s HBF with prefetching.
As the tile becomes narrower and taller, each 4\,KB request extends further beyond the compute tile boundary, increasing both read amplification and the number of outstanding requests per tile.
Performance degrades even when amplified data fits in the L2 cache, because the straggler effect depends on the number of outstanding requests.
The best performance is obtained when the memory tile fits entirely within the compute tile.

\begin{figure}[t]
    \centering
    \includegraphics[width=1.0\linewidth]{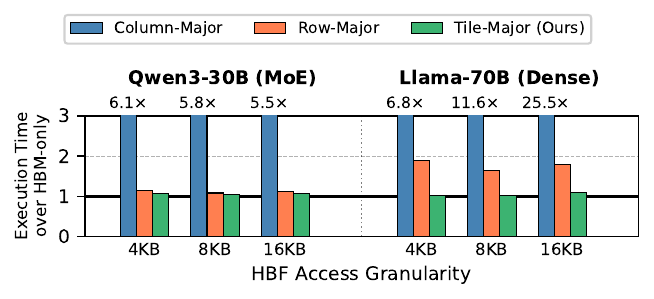}
    \vspace{-0.35in}
    \caption{HBF granularity sensitivity, normalized to HBM-only execution time. Tile-major remains effective at 8\,KB and 16\,KB access granularity.}
    \label{fig:sensitivity_granularity}
\end{figure}

Figure~\ref{fig:sensitivity_granularity} sweeps the HBF access granularity to 8\,KB and 16\,KB. Tile-major is insensitive to larger granularities, staying within 1.03--1.09$\times$ of the HBM baseline.
Column-major instead degrades sharply on Llama, from 6.8$\times$ to 25.5$\times$, as amplification scales with page size.

\subsection{\rev{Sensitivity: Split-K and Compute Tile Size}}
\label{sec:eval_kernel_parameter}

\begin{figure}[t]
    \centering
    \includegraphics[width=1.0\linewidth]{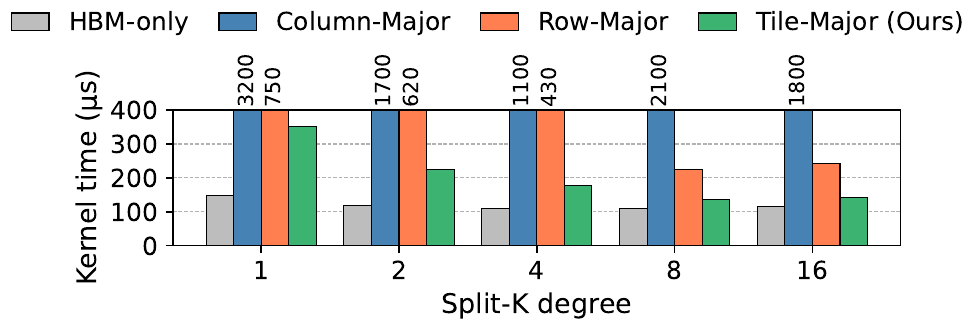}
    \vspace{-0.3in}
    \caption{\rev{Execution time of the Llama-70B down projection, vs.\ split-K degree at batch 16 on 10\,$\mu$s HBF. Each layout uses a prefetcher tuned for it.}}
    \label{fig:kernel_params}
\end{figure}

\rev{Tile-major remains the fastest layout across split-K degrees and compute tile sizes, including the largest tiles we evaluate.
The two parameters change together: split-K changes the number of CTAs, and the autotuner selects a different compute tile for each configuration.
Thus we sweep on both.

\parhead{Split-K.}
Without split-K, Llama's down projection launches only 128 CTAs at decode batch sizes.
Split-K factor $s$ runs $128s$ CTAs over contiguous $K/s$ slices.
More CTAs keep more HBF requests in flight, so both layouts speed up until $s=8$, the best degree for both.
At $s=8$, tile-major is 1.65$\times$ faster than row-major.
Tile-major comes within 1.24$\times$ of the best HBM-only time, while row-major stays at 2.05$\times$ (Figure~\ref{fig:kernel_params}).

\parhead{Compute tile size.}
With 256$\times$128 tiles on Qwen at batch 64, tile-major is 5.6\% slower than HBM-only, whereas column- and row-major are 11.05$\times$ and 2.37$\times$ slower.
Tile-major thus requires neither small compute tiles nor long $K$, since Qwen's experts have short $K$.
Even larger compute tiles are limited by shared memory capacity, leaving fewer resident CTAs and shallower pipelining.}

\subsection{\rev{Sensitivity: Prefetcher Aggressiveness}}
\label{sec:eval_prefetcher}

\begin{figure}[t]
    \centering
    \includegraphics[width=1.0\linewidth]{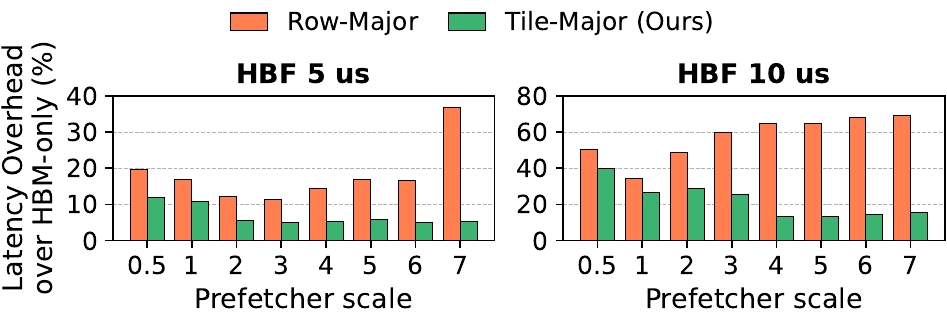}
    \vspace{-0.3in}
    \caption{\rev{Latency overhead of a Llama-70B decode over HBM-only across prefetcher scales (multipliers on the degree), at batch 16 on 5 and 10\,$\mu$s HBF.}}
    \label{fig:prefetch_degree}
\end{figure}

\rev{Figure~\ref{fig:prefetch_degree} multiplies the adaptive prefetch degree by a \emph{prefetcher scale} of 0.5 to 7 for both layouts.
Tile-major is faster than row-major at their respective best scale.
At its best scale, row-major is 11.3\% slower than HBM-only on 5\,$\mu$s HBF and 34.5\% slower on 10\,$\mu$s HBF, while tile-major is 5.0\% and 13.2\% slower.
Prefetching further ahead does not help row-major, because the performance of row-major relies on the fetched block being shared by multiple CTAs on L2, but prefetching too much leads to the memory blocks getting evicted before they are shared. 
On 10\,$\mu$s HBF, row-major's up/gate kernel slows from 210 to 498\,$\mu$s as the scale grows, while tile-major's stays between 212 and 238\,$\mu$s.
We note that row-major kernels are also autotuned for each batch size and layout, so the baseline uses its own best kernel configuration.}

\subsection{Read Amplification on RoMe}
\label{sec:eval_rome}

Finally, we evaluate tile-major layout on RoMe~\cite{nam2026rome}.
We assume RoMe increases the number of HBM3e channels from 16 to 18 per stack (12.5\% higher bandwidth), yielding 5.53\,TB/s total bandwidth across six stacks, with a 4\,KB access granularity.
Because RoMe latency remains at the scale of DRAM, no prefetcher is used.

\begin{figure}[t]
    \centering
    \includegraphics[width=1.0\linewidth]{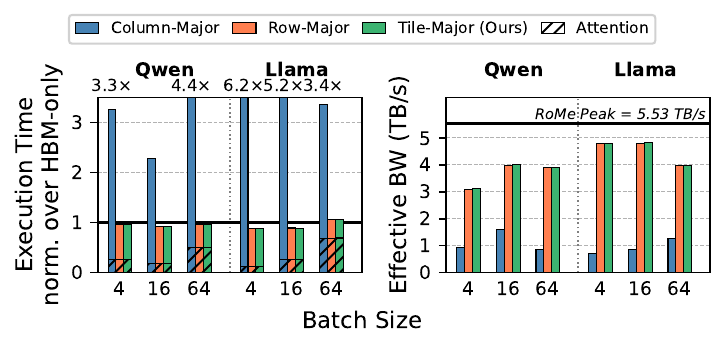}
    \vspace{-0.35in}
    \caption{\rev{Execution time and effective bandwidth on RoMe. Straggler effect disappears with low latency, but read amplification remains. Hatched segments show attention, which also reads the KV cache from RoMe.}}
    \label{fig:eval_rome}
\end{figure}

Figure~\ref{fig:eval_rome} shows execution time and effective bandwidth on a GPU with RoMe.
Column-major layout shows up to 6.2$\times$ slowdown due to bandwidth waste, confirming that the read amplification problem still exists in a DRAM with larger access granularity.
The straggler effect is not prominent in RoMe: the low latency of HBM can be hidden even with a smaller degree of memory-level parallelism, so row-major performs within 2\% of tile-major on a single GPU.

\rev{However, tensor parallelism (TP) worsens the straggler effect on RoMe.
With TP=8, Llama's O and down projections keep $N$ = 8,192 while $K$ shrinks by 8$\times$, and each launches only 128 CTAs, one per SM, so no other CTA can hide behind the row stragglers.
On these GEMMs, row-major is up to 19\% slower than tile-major.
Qwen3 with TP=4 shows no such gap, as its expert GEMMs still launch hundreds of CTAs that hide the row latency.
Under TP=8, the GEMM kernels tuned for column-major weights run up to 23\% faster than those tuned for row-major weights even on HBM, and tile-major retains this advantage on RoMe: it outperforms row-major by up to 28\% without split-K and 34\% with a split-K factor of 4.}

\rev{\parhead{KV cache and activation read/writes.}
Tile-major layout is unnecessary for the KV cache and activations, even on RoMe where they reside in LGMS.
Unlike matrix weights, the KV cache is a high-dimensional tensor (layer, token, head, dim) whose contiguous dimension is well under 4\,KB (e.g., 128 for Llama and Qwen).
Inference frameworks already support KV layouts in which a block of tokens of one head is contiguous (i.e., \texttt{LHBNC} layout~\cite{kwon2023efficient}), so a 16-token block forms a 4\,KB tile without remapping.
With such a layout, attention reads incur no read amplification except for the newest block being filled.
Activations are small (e.g., 2.8\,MB for Qwen3 at batch 64) and fit in L2, so their partial writes are merged before write-back.
Thus performance is minimally affected.
}

\section{Related Work} 
\label{sec:related}

\parhead{Address remapping for locality.}
Impulse~\cite{carter1999impulse}, Akin et al.~\cite{akin2015data}, and GS-DRAM~\cite{seshadri2015gather} remap addresses at the memory controller or inside the DRAM to improve spatial locality.
The works target cache-line-level locality on CPU DRAM, whereas tile-major targets 4\,KB page utilization on large-granularity GPU memory with no aliasing or cache modifications.

\parhead{Data layout optimization.}
Blocked layouts on CPUs~\cite{park2003tiling, chatterjee1999recursive}, XLA's tiled TPU arrays~\cite{openxla2026tiled}, and cuBLASLt's \texttt{COL32} layouts~\cite{nvidia2026cublas} store matrix tiles contiguously, but fix the tile shape to the cache line or compute unit and require dedicated kernels.
ASTA~\cite{sung2012dl}, Rhu et al.~\cite{rhu2013locality}, and CODA~\cite{kim2018coda} reorganize data layouts or adjust access granularity to improve coalescing and reduce over-fetching on standard GPU memory.
Tile-major instead targets the access granularity mismatch introduced by LGMS.

\parhead{Tiled memory layout in GPU texture units.}
GPU texture memory uses Morton/Z-order layouts to improve 2D spatial locality for 64--128\,B texture accesses~\cite{nvidia2026cuda}.
Tile-major instead targets 4\,KB pages with axis-aligned rectangular tiles, so each page belongs to a single compute tile, whereas bit-interleaved Morton addressing spans multiple tile boundaries.

\parhead{Large-granularity memory systems for LLM inference.}
H$^3$~\cite{ha2026h}, HAVEN~\cite{hsu2026haven}, and RoMe~\cite{nam2026rome} explore alternative GPU memory architectures, but none address read amplification of tiled matmuls under large access granularity.
ZnG~\cite{zhang2020zng}, BaM~\cite{qureshi2023gpu}, and LLM in a flash~\cite{alizadeh2024llm} target millisecond-latency SSD-class flash over PCIe, not co-packaged microsecond-latency HBF.
Recent works analyze throughput trade-offs of HBF~\cite{son2026exploring, yin2026potential, li2026hbf}, place high-reuse KV cache on HBF~\cite{kyung2026high}, and co-design the HBM+HBF hierarchy~\cite{park2026hbm, wang2026flashaccel, kim2026beyond}.
However, these studies rely on serving-level or analytical models rather than a microarchitectural performance model.
To the best of our knowledge, this work is the first to analyze and address the read amplification problem on HBF-based GPUs.

\parhead{GPU prefetching.}
Prior GPU prefetching work focuses on \emph{what} to prefetch in irregular DRAM-backed workloads~\cite{lee2010many,jog2013orchestrated,koo2018cta}.
With tile-major layout, accesses become deterministic, so the problem shifts to \emph{how aggressively} to prefetch.

\section{Conclusion}
Large-granularity memory systems promise the capacity and bandwidth that LLM inference demands, but their kilobyte-scale accesses clash with the 2-D tiles of GPU matmuls, causing severe read amplification.
\emph{Tile-major layout} reshapes each 4\,KB region into a rectangle that fits within a compute tile, and \emph{\papername} makes this layout transparent to existing kernels through lightweight DSL and TMA extensions.
With tile-major and adaptive prefetching, an HBF-augmented GPU goes from a 1.62--6.69$\times$ slowdown to within 10\% of an HBM-only baseline at 5\,$\mu$s NAND latency.


\bibliographystyle{IEEEtranS}
\bibliography{refs}

\end{document}